\documentclass[prd,twocolumn,showpacs,nofootinbib,amsmath,amssymb,superscriptaddress]{revtex4-1}

\usepackage{graphicx}
\usepackage{dcolumn}
\usepackage{bm}
\usepackage{relsize}
\usepackage{amsmath,amssymb,bbm,mathrsfs,nicefrac}
 \usepackage{array,multirow}
 \usepackage{float}
\usepackage{tikz}
\usepackage{amsmath}
\usepackage{amssymb}
\usepackage{graphicx,textcomp,float,gensymb,wrapfig, enumitem,comment,dsfont,framed,slashed,appendix,wrapfig,xcolor}
\usepackage{ bbold }
\usepackage{braket} 
\usepackage{ mathrsfs }
\usepackage{etoolbox,hyperref,makecell}
\usepackage{feynmp}
\usepackage[normalem]{ulem}
\usetikzlibrary{arrows.meta}
\usepackage{empheq}
\usepackage[normalem]{ulem}
\usepackage{scrextend}
\usepackage{slashed}
\newcommand{\bea}{\begin{eqnarray}}  
\newcommand{\eea}{\end{eqnarray}}

\definecolor{darkgreen}{rgb}{0.2, 0.3, 0.1}

\newcommand{\beq}{\begin{equation}}
\newcommand{\eeq}{\end{equation}}

\newcommand{\calO}{{\cal O}}

\newcommand{\MeV}{\, {\rm MeV}}
\newcommand{\fm}{\, {\rm fm}}

\usepackage[utf8]{inputenc}

\begin{document}

\title{
Critical lensing and kurtosis near a critical point in the QCD phase diagram in and out-of-equilibrium
}

\author{Travis Dore}
\email[Corresponding author: ]{tdore2@illinois.edu}
\affiliation{Illinois Center for Advanced Studies of the Universe, Department of Physics, University of Illinois at Urbana-Champaign, Urbana, IL 61801, USA}

\author{Jamie M. Karthein}
\affiliation{Center for Theoretical Physics, Massachusetts Institute of Technology, Cambridge, MA 02139, USA}

\author{Isaac Long}
\affiliation{Illinois Center for Advanced Studies of the Universe, Department of Physics, University of Illinois at Urbana-Champaign, Urbana, IL 61801, USA}

\author{Debora Mroczek}
%\email{deboram2@illinois.edu}
\affiliation{Illinois Center for Advanced Studies of the Universe, Department of Physics, University of Illinois at Urbana-Champaign, Urbana, IL 61801, USA}

\author{Jacquelyn Noronha-Hostler}
%\email{jnorhos@illinois.edu}
\affiliation{Illinois Center for Advanced Studies of the Universe, Department of Physics, University of Illinois at Urbana-Champaign, Urbana, IL 61801, USA}

\author{Paolo Parotto}
\affiliation{Pennsylvania State University, Department of Physics, University Park, PA 16802, USA}

\author{Claudia Ratti}
\affiliation{Physics Department, University of Houston, Houston, TX 77204, USA }

\author{Yukari Yamauchi}
%\email{yyukari@umd.edu}
\affiliation{Department of Physics, University of Maryland, College Park, MD 20742, USA}

\date{\today}

\begin{abstract}
 
In this work, we study the lensing effect of the QCD critical point on hydrodynamic trajectories, and its consequences on the  
net-proton kurtosis $\kappa_4$. Including critical behavior by means of the BEST Collaboration equation of state (EoS), we first consider a scenario in equilibrium, then compare with hydrodynamic 0+1D simulations with Bjorken expansion, including both shear and bulk viscous terms. We find that, both in and out-of-equilibrium, the size and shape of the critical region directly affect if the signal will survive through the dynamical evolution.

\end{abstract}

\maketitle

\section{Introduction}

Understanding the phase structure of Quantum Chromodynamic matter has been one of the major endeavors in nuclear physics for the past several decades. While it is well understood that a cross-over transition from the Quark Gluon Plasma (QGP) into a hadron resonance gas exists at vanishing baryon densities \cite{Aoki:2006we, Bhattacharya:2014ara,Bazavov:2018mes,Borsanyi:2010bp,Borsanyi:2020fev}, it is conjectured that at very large densities a first-order phase transition should appear \cite{Stephanov:1998dy}. In that case, a critical point would exist at the boundary between the cross-over and first-order phase transitions. At the critical point, the transition would be of second-order. 

Due to the fermion sign problem, it is not possible to calculate the QCD equation of state directly at finite baryon densities with lattice QCD simulations, and thus locate the critical point \cite{Troyer:2004ge,Ratti:2018ksb}. Therefore, its existence and location have not yet been confirmed. On the other hand, a number of effective models that reproduce lattice QCD results at low baryon densities predict a critical point at large baryon chemical potentials \cite{Halasz:1998qr,Stephanov:1998dy,Stephanov:1999zu,Ratti:2006gh,Dexheimer:2009hi,Eichmann:2015kfa,Critelli:2017oub,Fan:2016ovc,Fu:2019hdw,Motornenko:2019arp,Annala:2019puf,Tan:2020ics,Grefa:2021qvt,Gunkel:2021oya,Grefa:2022sav}. The critical point might be reachable within low-energy heavy-ion collisions at accelerators such as the Relativistic Heavy-Ion Collider (RHIC) as well as future facilities such as the Facility for Antiproton and Ion Research (FAIR) \cite{Bzdak:2019pkr}.

At the moment, the primary signature of the critical point is a peak in the kurtosis $\kappa_4$ of measured net-proton distributions \cite{Stephanov:2008qz, Athanasiou:2010kw}. From the theoretical point of view, one defines the susceptibilities of baryon number as $\chi_n \equiv \partial^n (p/T^4)/\partial (\mu_B/T)^n$, where $p$ is the QCD pressure. It is possible to relate the kurtosis to the susceptibilities as follows: $\kappa_4\sigma^2=\chi_4/\chi_2$, where $\sigma^2$ is the variance of the net-proton distribution. This relationship is not strictly exact, since the measured kurtosis is for net-protons, while the theoretical quantity relates to net-baryon number \cite{Kitazawa:2011wh,Kitazawa:2012at,Nahrgang:2014fza,Vovchenko:2017ayq}. Right at the critical point one expects a divergence in $\kappa_4$, because it scales with the correlation length $\xi$ as $\kappa_4 \propto \xi^7$ \cite{Stephanov:2011pb}. The higher the order of the susceptibility, the larger the power of $\xi$ it scales with. For this reason, higher order moments are the observables of choice for the detection of the critical point, with the kurtosis being (currently) the best compromise in terms of 
signal to noise ratio in experiments.

The qualitative features of the kurtosis $\kappa_4$ have been previously studied in the context of a mapping of critical behavior in the 3D Ising model onto the QCD phase diagram, both without \cite{Stephanov:2011pb} and with \cite{Mroczek:2020rpm} the inclusion of all sub-leading terms in the vicinity of the critical point. In the latter case, it was shown that the specifics of the Ising-to-QCD mapping have a strong influence on the resulting shape of the critical region, and in turn on the height and width of $\kappa_4 $ at freeze-out. The behavior of the net-baryon kurtosis at finite density was also studied in other approaches, see e.g. \cite{Critelli:2017oub,Vovchenko:2017ayq,Fu:2021oaw,Vovchenko:2022szk}.

However, previous studies of the kurtosis focused on equilibrium properties, whereas it is well-known that the QGP is probed dynamically in heavy-ion collisions, flowing like a relativistic viscous fluid \cite{Danielewicz:1984ww,Kovtun:2004de,Romatschke:2007mq,Bozek:2011ua,Heinz:2013th,Luzum:2013yya,Niemi:2015qia,Noronha-Hostler:2015uye,McDonald:2016vlt,Bernhard:2019bmu,Alba:2017hhe}. In fact, some studies suggest that the shear-viscosity-over-enthalpy ratio $\eta T/w$ increases significantly at large baryon densities \cite{Kadam:2014cua,Auvinen:2017fjw,Soloveva:2020hpr,McLaughlin:2021dph} (although critical scaling for $\eta T/w$ appears to be negligible \cite{Monnai:2016kud}). Even more importantly, the bulk viscosity $\zeta$ increases when the speed of sound $c_s^2$ approaches $c_s^2\rightarrow 0$ (as it does at the critical point, where this behavior is further affected by critical scaling \cite{Abbasi:2021rlp}). Thus, a peak in $\zeta T/w$ at the critical point is expected \cite{Dore:2020jye}, which is also further enhanced due to criticality, as the bulk viscosity itself scales with $\zeta \propto \xi^3$ \cite{Monnai:2016kud,Martinez:2019bsn,Rajagopal:2019xwg,Dore:2020jye}. 

Recent studies have probed the applicability of hydrodynamics near the QCD critical point and potential dynamical signatures of the critical point \cite{Monnai:2016kud,Rougemont:2018ivt,Critelli:2018osu,Rajagopal:2019xwg,Dore:2020jye,Monnai:2021kgu,An:2020vri,Du:2020bxp,An:2021wof,Pradeep:2022mkf}. Critical points can deform ideal hydrodynamics trajectories, causing them to merge towards the critical point \cite{Stephanov:1998dy,Nonaka:2004pg,Asakawa:2008ti,Grefa:2021qvt,Karthein:2021nxe}. This effect is known as critical lensing \cite{Stephanov:2004wx,Nonaka:2004pg,Asakawa:2008ti}.  However, it was found recently that far-from-equilibrium effects at a critical point \cite{Dore:2020jye} or first-order phase transition \cite{Feng:2018anl} can also dramatically alter the path through the QCD phase diagram, a fact confirmed also by later works \cite{Du:2020zqg,Du:2021zqz}. Thus, it is not clear what interplay exists between critical lensing and viscous effects. Furthermore, a connection has not yet been made between the size and shape of the critical region itself and potential signatures of criticality. 
The question naturally arises: can far-from-equilibrium hydrodynamics smear out any potential signs of the critical point?  

Currently, a full, dynamical framework does not exist to properly describe the evolution of a system in the vicinity of the critical point. Realistically, one would require an event-by-event analysis with 3+1D relativistic viscous hydrodynamics with BSQ (baryon number, strangeness, and electric charge) conserved charges and critical fluctuations  (see \cite{Rao:2019vgy,Dexheimer:2020zzs,An:2021wof} for more details).  While significant efforts have been made in this direction \cite{Karpenko:2013wva,Rougemont:2015ona,Stephanov:2017ghc,Feng:2018anl,Nahrgang:2018afz,Du:2019obx,Denicol:2018wdp,Batyuk:2017sku,Fotakis:2019nbq}, the community is still a long way from reaching this milestone.  In the meantime, it is useful to obtain qualitative understanding from simplified models to guide experiments and future theoretical studies, once dynamical models improve over time.

In this work, we explore the lensing effect of the critical point on evolution trajectories, and its implications on the kurtosis of net-proton number distributions. First, we do so in an equilibrium scenario, by incorporating critical behavior through the BEST Collaboration EoS \cite{Parotto:2018pwx}. Secondly, we study the effect of out-of-equilibrium physics by means of simple 0+1D hydrodynamic simulations with Bjorken expansion, which include both shear and bulk viscosities \cite{Dore:2020jye}. In both cases, we investigate how the non-universal parameters of the Ising-to-QCD map of the BEST EoS, which have been shown to determine the size and shape of the critical region \cite{Mroczek:2020rpm}, also influence the lensing effect and the resulting net-proton kurtosis.

We find that, in the cases where the critical region extends predominantly in the temperature direction, critical lensing is enhanced, via a clustering of evolution trajectories around the critical point, both in and out-of-equilibrium. In contrast, when the critical region predominantly extends in the $\mu_B$ direction, the effect is significantly weaker and very few hydrodynamic trajectories deviate towards the critical point. In general, we find that both viscous effects and the shape of the critical region are crucial to the discussion of critical lensing. Due to the intriguing results presented in this work, future plans are already underway to explore these effects in higher dimensions, and in a framework that incorporates BSQ diffusion.

\section{Model }

\subsection{Equation of State} \label{sec:EoS}

In this work, we incorporate the effect of a critical point primarily through the equation of state. We use the procedure, and the notation, developed in  Ref.~\cite{Parotto:2018pwx} for constructing a family of EoS with a critical point. By construction, these EoSs match lattice QCD results at $\mu_B=0$ up to order $\calO (\mu_B^4)$,
and contain a critical point in the 3D Ising model universality class. 

The procedure is based on a parametrization of the 3D Ising model EoS in the vicinity of the critical point \cite{Nonaka:2004pg,Guida:1996ep,Schofield:1969zz,Bluhm:2006av}, and a subsequent mapping of 3D Ising variables (reduced temperature $r = (T - T_c) / T_c$ and magnetic field $h$) to QCD variables, temperature $T$ and baryon chemical potential $\mu_B$. 
We follow Ref.~\cite{Parotto:2018pwx}, which implements a linear map
\cite{Rehr:1973zz}: 
\begin{align}  
\frac{T - T_C}{T_C} &=  w \left( r \rho \,  \sin \alpha_1  + h \, \sin \alpha_2 \right) \, \, , \label{eq:IsQCDmap1} \\ \nonumber
\frac{\mu_B - \mu_{BC}}{T_C} &=  w \left( - r \rho \, \cos \alpha_1 - h \, \cos \alpha_2 \right) \, \, , \label{eq:IsQCDmap2}
\end{align} 
where $(T_C,\mu_{BC})$ indicate the location of the critical point, and $(\alpha_1, \alpha_2)$ are the angles between the horizontal $(T=\textit{const})$ lines and the $h=0$ and $r=0$ Ising model axes, respectively. Finally, $w, \rho$ are scaling parameters, with $w$ 
determining the global scaling of both $r$ and $h$, and $\rho$ determining the relative scaling between the two.

While such a linear map contains six parameters, it is possible to reduce them to four, as was done in \cite{Parotto:2018pwx}, by imposing that the critical point lies on the chiral transition line predicted by lattice QCD \cite{Bellwied:2015rza}:
\begin{equation}\label{eq:trline}
T = T_0 + \kappa_2 \, T_0 \left( \frac{\mu_B}{T_0} \right)^2 + {\cal O} (\mu_B^4),
\end{equation}
from which one can obtain $T_C$ and $\alpha_1$, given a value of $\mu_{BC}$. As in the original formulation, we use $\kappa_2 = -0.0149$ from Ref.~\cite{Bellwied:2015rza}. This value is consistent with more recent results, which also predict the next-to-leading order coefficient to vanish within error bars \cite{Bazavov:2018mes,Borsanyi:2020fev}. 

Exact matching to lattice QCD at $\mu_B=0$ is imposed by requiring that the Taylor coefficients used in the expansion of the pressure obey
\begin{equation} \label{eq:coeffs}
T^4 c_n^{\text{LAT}} (T) = T^4 c_n^{\text{Non-Ising}} (T) + T_C^4 c_n^{\text{Ising}} (T) \, \, ,
\end{equation}
where $c_n^{\text{LAT}}$ are lattice Taylor QCD coefficients ~\cite{Borsanyi:2013bia, Bellwied:2015lba}. Here, the $c_n^{\text{Ising}}$ determine the contribution to the lattice coefficients due to the presence of the critical point, and the $c_n^{\text{Non-Ising}}$ are \textit{defined} as the contribution at vanishing $\mu_B$ from a non-critical background field, namely as the difference between the lattice and Ising coefficients.

The full pressure is reconstructed as 
\begin{equation} \label{eq:Pfull}
P (T, \mu_B) = T^4 \sum_n c_n^{\text{Non-Ising}} (T) \left( \frac{\mu_B}{T} \right)^n + P^{\text{QCD}}_{\text{crit}}(T, \mu_B) \, \, ,
\end{equation}
where $P^{\text{QCD}}_{\text{crit}}(T, \mu_B)$ is the critical pressure mapped onto QCD from the 3D Ising model, which has been symmetrized about $\mu_B=0$. The full EoS is then derived from Eq. (\ref{eq:Pfull}) via standard thermodynamics relations.

With this procedure, each realization of the equation of state varies based on the \textit{non-universal} mapping of Eq.\eqref{eq:IsQCDmap1}, thus on the parameters $\mu_{BC}, w, \ \rho, \ \textrm{and} \ \Delta\alpha = \alpha_2 - \alpha_1$. For additional details, we refer the reader to Ref.~\cite{Parotto:2018pwx}. This scheme was recently expanded to include the correct charge conservation constraints for ultra-relativistic heavy-ion collisions (see Ref. \cite{Karthein:2021nxe}). In this work, we assume $\mu_S = \mu_Q = 0$, as in the original framework.

Finally, the correlation length is also calculated within the BEST collaboration code as in Ref. \cite{Karthein:2021nxe}. It follows Widom’s scaling form in terms of Ising model variables as shown in  Refs. \cite{Brezin:1976pt, Berdnikov:1999ph, Nonaka:2004pg}: 

\begin{equation} \label{corr_length}
    \begin{split}
        \xi^2(r,M) = f^2 |M|^{-2 \nu / \beta} g(x),
    \end{split}
\end{equation}
where $f$ is a constant with the dimension of length, which we set to 1 fm, $\nu$ = 0.63 is the correlation length critical exponent in the 3D Ising Model, $g(x)$ is the scaling function and the scaling parameter is $x$=$\frac{|r|}{|M|^{1/\beta}}$. For further details, we refer the reader to Ref. \cite{Karthein:2021nxe}.

\subsection{Hydrodynamic Setup} 
\label{hydro_Setup}

The correct relativistic hydrodynamic description of a system in the vicinity of a critical point is still an open question. As far as critical fluctuations of the critical mode are concerned, progress has been made in recent years \cite{Son:2004iv,Stephanov:2017ghc,Akamatsu:2018vjr,An:2019csj,Young:2014pka,Sakai:2017rfi,Singh:2018dpk,Du:2019obx}.  However, a clear consensus has not yet emerged. We do not include fluctuations of the critical mode in this work. We also do not include effects from Kibble-Zurek scaling \cite{Kibble:1980mv,Zurek1985} which also may be relevant in the critical region during the transition \cite{Mukherjee:2016kyu}. Nonetheless, we remain sensitive to critical behavior both through the equation of state, and through the critical scaling of the bulk viscosity.

The hydrodynamic setup of the current work is the same as that of Ref.\ \cite{Dore:2020jye}, where more details can be found. In order to qualitatively investigate the influence of out-of-equilibrium initial conditions and different EoS on hydrodynamic trajectories in the QCD phase diagram, as well as on potential observables, we employ the highly symmetric Bjorken flow picture. While the symmetry constraints of Bjorken flow are no longer understood to be good approximations at lower beam energies, they can certainly provide valuable intuition on the response of the hydrodynamic system to different EoS.

The equations of motion used in this work are based on the idea that the dissipative currents, such as the shear-stress tensor $\pi^{\mu\nu}$ and bulk scalar $\Pi$, evolve according to relaxation equations that describe how such quantities deviate from their relativistic Navier-Stokes values, which is required for any relativistic viscous hydrodynamic equations to ensure causality and stability. There are three different methods for relativistic viscous fluids: phenomenological Israel-Stewart \cite{Israel:1979wp}, DNMR \cite{Denicol:2012cn}, and BDNK \cite{Bemfica:2017wps,Bemfica:2020zjp,Kovtun:2019hdm,Hoult:2020eho}. 
In \cite{Dore:2020jye}, phenomenological Israel-Stewart and DNMR equations of motion were compared at a critical point and it was found that DNMR are more well-behaved numerically when traversing the critical region with a critically scaled bulk viscosity. Due to the fact that the BDNK equations of motion are more recent, they have yet to be checked at this time for a non-conformal EOS (see \cite{Pandya:2022pif}). Thus, we will only focus on DNMR equations of motion for this study. Using hyperbolic coordinates with the metric $g_{\mu\nu} = \text{diag}(1,-1,-1,-\tau^2)$, the underlying symmetries of Bjorken flow imply that all dynamical quantities depend only on the proper time $\tau = \sqrt{t^2-z^2}$ and the equations reduce to \cite{Denicol:2012cn,Bazow:2016yra} \begin{align}
    \dot{\varepsilon}&=-\frac{1}{\tau}\left[\varepsilon+p+\Pi-\pi^{\eta}_{\eta}\right]\\
      \tau_{\pi}\dot{\pi}^{\eta}_{\eta}+\pi^{\eta}_{\eta}&=\frac{1}{\tau}\left[\frac{4\eta}{3} - \pi^\eta_\eta\left(\delta_{\pi\pi} + \tau_{\pi\pi} \right) + \lambda_{\pi\Pi}\Pi \right]\label{eq:shearEvo}\\
 \tau_{\Pi}\dot{\Pi}+\Pi&=-\frac{1}{\tau}\left(\zeta + \delta_{\Pi\Pi}\Pi+\frac{2}{3}\lambda_{\Pi\pi}\pi^{\eta}_{\eta}\right)\label{eq:bulkEvo}\\
    \dot n &= -\frac{n}{\tau}\label{eqn:rhoBevo}.
\end{align}
We note that, in Bjorken flow, the particle diffusion contribution vanishes and, thus, the baryon density equation can be readily solved to give $n(\tau)=n_0 (\tau_0/\tau)$, where $n_0$ and $\tau_0$ are the initial baryon density and time, respectively. The definitions of second order transport coefficients and the functional dependence of shear viscosity on $T$ and $\mu_B$ can be found in \cite{Dore:2020jye,McLaughlin:2021dph}. 

We would like to emphasize the importance of Eqs.\ (\ref{eq:shearEvo}),(\ref{eq:bulkEvo}) in our analysis. Since the shear and bulk viscous terms are dynamical quantities, they require their own initial conditions, which allows us to explore many different hydrodynamic trajectories through the phase diagram. After initializing the system at different densities with different initial conditions in the shear and bulk sectors, we select hydrodynamic trajectories that traverse the critical region. In practice, we select trajectories that pass, along lines parallel to the chiral transition line and shifted downwards by an amount $\Delta T$, within a width of $3.5$ MeV on either side of the ideal hydrodynamic trajectory (i.e. the isentropic trajectory) that passes through the critical point. This means that the trajectories we select populate a total width of $7$ MeV, with the isentrope at the center.
The initial conditions of the system are constrained by the Weak Energy Condition \cite{Janik:2005zt, Dore:2020jye} which allows us to initialize the system with
\begin{equation*}
    \left\{\frac{\pi^\eta_\eta}{\varepsilon + p},\frac{\Pi}{\varepsilon + p}\right\}_0 \equiv \{\chi,\Omega\}_0 \in [-0.5,0.5].
\end{equation*}
In this paper, when we plot hydrodynamic trajectories we will use a specific color scheme depending on their respective initial conditions, which was  explained in Fig.\ 2 from \cite{Dore:2020jye}.  Notice that the purple, blue, and turquoise lines indicate initial conditions that are consistent with those typically found from heavy-ion collisions where  $\Pi<0$ and $\pi^\eta_\eta>0$.  In contrast, the red and orange lines indicate initial conditions where $\Pi>0$ and $\pi^\eta_\eta<0$, which are atypical for heavy-ion collisions. Additional details will be discussed in Sec.\ \ref{sec:hydroTraj}. 

The bulk viscosity used in this work is also the same as that in Ref.\ \cite{Dore:2020jye}. The expression for the critically scaled bulk viscosity is 
 \begin{equation}\label{eqn:zetaCS}
     \left(\frac{\zeta T}{w}\right)_{CS} =\frac{\zeta T}{w}
     \left[1 +\left(\frac{\xi}{\xi_0}\right)^3\right],
 \end{equation}
as  has been used in previous works  \cite{Monnai:2016kud,Martinez:2019bsn,Rajagopal:2019xwg,Dore:2020jye}. This ensures finite bulk viscosity outside the critical region, which is relevant for our work, as it influences the system's approach to the critical point. The shear viscosity $\eta T/w(T,\mu_B)$ that we employ, comes from the phenomenological approach in  Ref.~\cite{McLaughlin:2021dph}, where a hadron resonance gas model at low $T$ was matched to a functional form for the QGP. This model ensures that the minimum of $\eta T/w(T,\mu_B)$ will pass through the critical point and then follow along the first-order line. 

\section{Observables}

\subsection{Kurtosis} 

The kurtosis of net-baryon number distributions is currently, as mentioned, the most promising signature for a potential experimental detection of the QCD critical point in heavy-ion collisions. In practice, it can be directly connected \cite{Kitazawa:2011wh,Kitazawa:2012at,Nahrgang:2014fza} to the fluctuations of the net-proton ($N_p$) distribution that appear on an event-by-event basis and can be measured.  
Most experiments, including STAR \cite{STAR:2020tga}, HADES \cite{HADES:2020wpc}, and ALICE \cite{ALICE:2019nbs} measure the cumulants $\kappa_n$, of the net-p distribution, which are defined as:
\begin{center}
\begin{tabular}{lcl}
   mean & \hspace{0.2\linewidth} & $M$ = $\kappa_1 = M_1$ \\
   variance  & \quad & $\sigma^2$ = $\kappa_2 = M_2$\\
   skewness & \quad & $S\,\,$ = $\kappa_3=M_3$ \\
   kurtosis  & \quad & $\kappa_4=M_4-3M_2^2$
\end{tabular}
\end{center}
where $M_n$ is the $n^{th}$ moment of the distribution. 

Note that these measurements are beholden to the acceptance cuts of the detector.  Significant amount of effort has been made to increase the available rapidity window, because this has been shown to push the kurtosis measurements closer to the equilibrium values \cite{STARnote,Bzdak:2017ltv}.  However, a careful reader might also realize that if all particles were measured and used to calculate net-charge fluctuations, the results would be trivial.  For instance, heavy-ion collisions must always have global strangeness neutrality since strangeness is conserved and the initial ions do not carry any net-$S$.  Thus, for full acceptance net-$S$=0.  Similarly, for net-p and full acceptance, the only information provided would be the number of baryons stopped in the initial state and how that number fluctuates for a fixed centrality class.  Thus, there is an optimal window between too small vs. too large kinematic cuts that can yield the actual fluctuations of a net-charge that are sensitive to long range correlations \cite{Jeon:2000wg,Koch:2008ia}. Other factors that might impact the experimental measurements of fluctuations include canonical ensemble effects \cite{Bzdak:2012an,Vovchenko:2020tsr,Vovchenko:2020gne,Braun-Munzinger:2020jbk,Vovchenko:2021kxx}, coordinate vs momentum space \cite{Ling:2015yau,Ohnishi:2016bdf}, volume fluctuations \cite{Gorenstein:2011vq,Skokov:2012ds,Luo:2013bmi,Braun-Munzinger:2016yjz}, interactions in the hadronic phase \cite{Steinheimer:2016cir,Savchuk:2021aog}, and non-equilibrium effects \cite{Mukherjee:2015swa,Asakawa:2019kek}.

Keeping these caveats in mind, the cumulants defined above can be related to the so-called susceptibilities of baryon number:
\begin{equation} \label{eq:suscB}
    \chi^{B}_{n}=\frac{\partial^{n} p(\mu_B,T)/T^4}{\partial \left(\mu_B/T\right)^n}
\end{equation}
which can usually be calculated straightforwardly from theory, as they require simple derivatives of the pressure. In terms of the latter, the cumulants $\kappa_n$ read:
\begin{eqnarray}
M &=&\chi_1\\
\sigma^2 &=&\chi_2\\
S &=&\frac{\chi_3}{\chi_2^{3/2}}\\
\kappa_4 &=& \frac{\chi_4}{\chi_2^2}
\end{eqnarray}

Since the susceptibilities are \emph{extensive} variables that depend -- linearly, in a homogeneous system -- on the volume of the system, it is common to define ratios whose (leading) volume dependence is removed:
\begin{eqnarray}
    \frac{M}{\sigma^2}&=&\frac{\chi_1}{\chi_2}\label{eqn:chi12}\\
    S \sigma&=&\frac{\chi_3}{\chi_2}\\
    S\sigma^3 M^{-1}&=&\frac{\chi_3}{\chi_1}\\
    \kappa_4 \sigma^2&=&\frac{\chi_4}{\chi_2}\label{eqn:chi42}.
\end{eqnarray}

Most commonly, experimental results are shown for the ratios in Eqs. (\ref{eqn:chi12}-\ref{eqn:chi42}), because of the aforementioned advantage of removing the leading volume dependence. Hence, what is often referred to as ``kurtosis'', e.g. in the well-known plot from \cite{STAR:2020tga}, 
is indeed $\kappa_4 \sigma^2$.
Since net-proton fluctuations are the most sensitive to criticality, we consider only the baryon susceptibilities from now on and assume that protons are a good proxy for net-baryons.
If one is specifically interested in the interplay between all three conserved charges, better proxies could be used \cite{Bellwied:2019pxh}.  However, in our current set-up we cannot distinguish between protons and neutrons and, therefore, leave this issue for a later work.

When comparing net-particle fluctuations from theory and experiment with the aim of studying bulk thermodynamic properties of the system, it is customary to focus on central collisions, because they contain the largest number of participants, and are then more likely to be close to equilibrium. 

It is now well-understood in the hydrodynamic community that even central collisions in large systems initially begin far-from-equilibrium \cite{Noronha-Hostler:2015coa,Schenke:2019pmk,Summerfield:2021oex,Chiu:2021muk,Plumberg:2021bme}. While in idealized systems (0+1D Bjoerken flow with only shear viscosity) with a trivial EoS ($\varepsilon=3p$) it appears that  universal attractors appear \cite{Heller:2015dha,Buchel:2016cbj,Heller:2016rtz,Spalinski:2017mel,Romatschke:2017acs,Romatschke:2017vte,Behtash:2017wqg,Strickland:2017kux,Denicol:2017lxn,Blaizot:2017ucy,Casalderrey-Solana:2017zyh,Florkowski:2017olj,Heller:2018qvh,Rougemont:2018ivt,Denicol:2018pak,Almaalol:2018ynx,Casalderrey-Solana:2018uag,Behtash:2018moe,Behtash:2019txb,Strickland:2018ayk,Kurkela:2018wud,Strickland:2019hff,Kurkela:2019set,Jaiswal:2019cju,Denicol:2019lio,Brewer:2019oha,Almaalol:2020rnu,Berges:2020fwq,Bemfica:2020xym,NunesdaSilva:2020bfs,Attems:2017zam,Attems:2018gou}, the use of a non-trivial EoS and the inclusion of bulk viscosity significantly complicate the picture \cite{Romatschke:2017acs,Dore:2020jye,Chattopadhyay:2021ive}. 

In fact, near the QCD critical point there may not be enough time for a universal attractor to be reached \cite{Dore:2020jye}, also because the hydrodynamic phase appears to be significantly shorter at lower beam energies \cite{Auvinen:2013sba}. This means that some memory of the initial conditions is retained by the system until the final stages, and far-from-equilibrium effects will be crucial for understanding the influence of initial conditions on kurtosis measurements.

Finally, an additional complication is the discrepancy between the temperature and chemical potential at which hadrons are formed (i.e. the critical temperature and chemical potential $\left\{T_C,\mu_{BC}\right\}$) and the temperature and chemical potential $\left\{T_{FO},\mu_{B,FO}\right\}$ at which chemical freeze-out occurs. The chemical freeze-out is the stage in the evolution of the system at which inelastic collisions between hadrons cease, and particle multiplicities can usually be well-described by a hadron resonance gas. However, it is likely that $\left\{T_{FO},\mu_{B,FO}\right\}$ is not significantly below $\left\{T_C,\mu_{BC}\right\}$ because the hadron resonance gas produces many short-lived, heavy resonances that quickly push the system into chemical equilibrium \cite{Wong:1996va,Rapp:2000gy,Greiner:2000tu,Greiner:2004vm,NoronhaHostler:2007jf,NoronhaHostler:2009cf,Noronha-Hostler:2014aia,Almasi:2014pya,Beitel:2016ghw,Gallmeister:2017ths}. In order to take this uncertainty into account, we will consider three different scenarios in which $\Delta T=T_C-T_{FO} = 1,3,5 \MeV$.
 
\subsection{Critical Lensing}\label{sec:criticallensing}

\begin{figure*}[t]
    \centering
    \begin{tabular}{c c}
    \includegraphics[width=.45\linewidth]{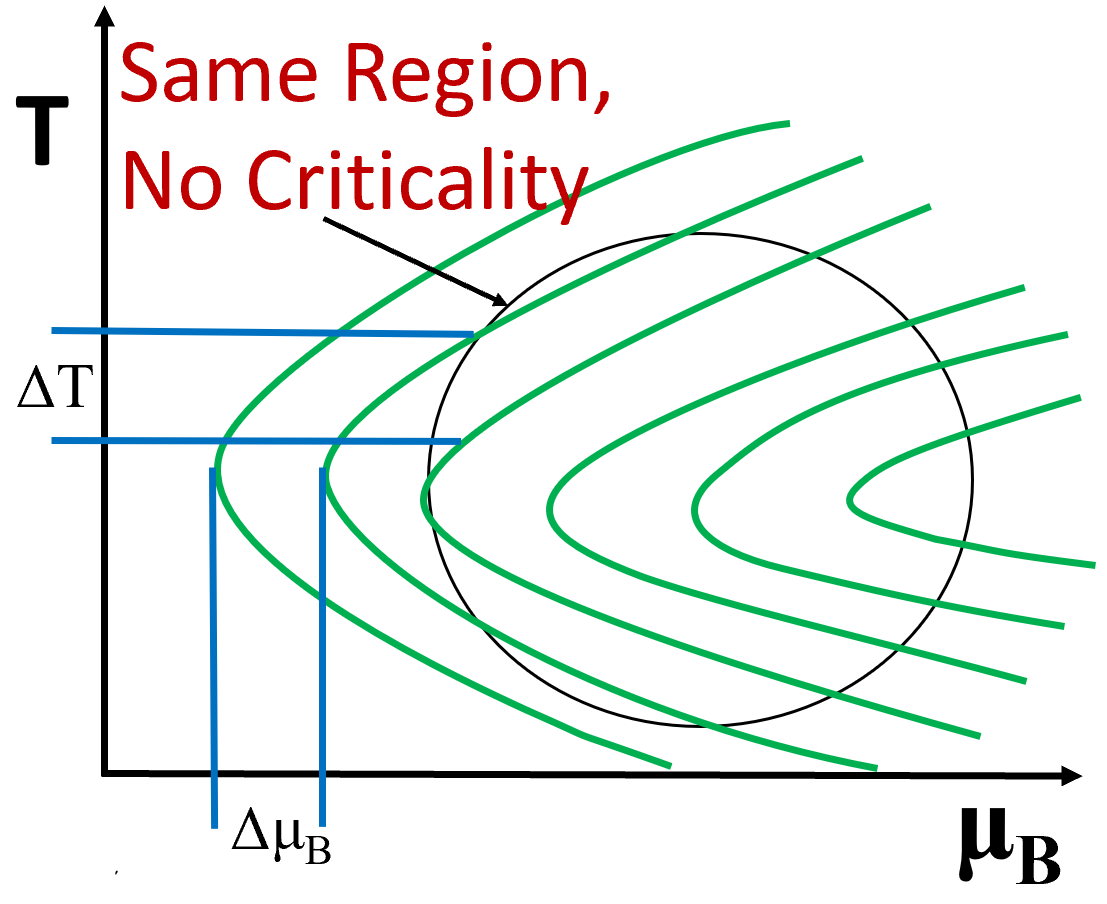} &
    \includegraphics[width=.45\linewidth]{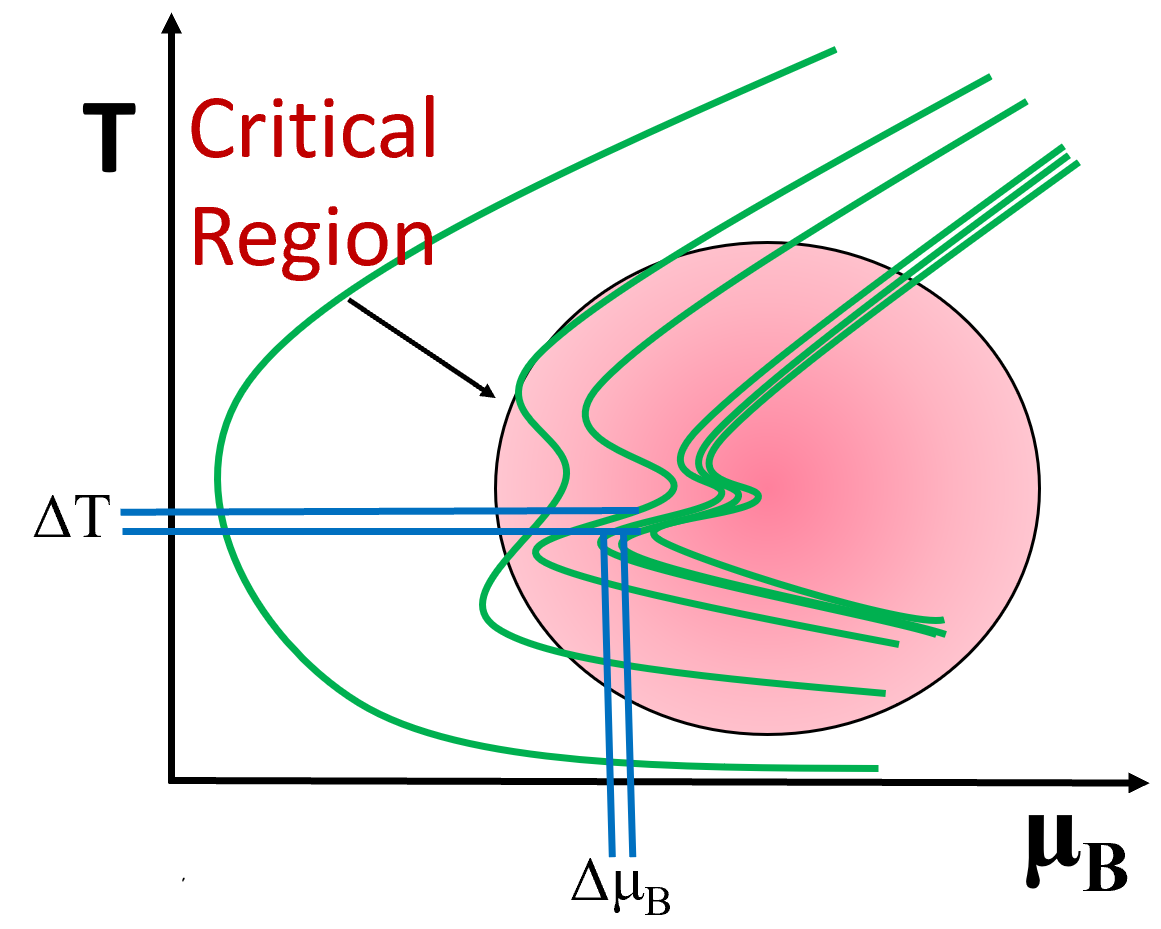} 
    \end{tabular}
    \caption{Schematic visualization of critical lensing, through a comparison of isentropic trajectories with (right) and without (left) criticality. Both circular regions are centered around the same value, and are the same size. It can be shown (see main text) that the spacing between curves, $(\Delta T, \Delta \mu_B)$, is smaller in the critical case. This leads to a larger density of trajectories crossing through the same region, given the same set of initial conditions.}
    \label{fig:schem_traj}
\end{figure*}

An interesting question regarding the effect of a critical point in the QCD phase diagram is, to what extent it can affect hydrodynamic evolution trajectories, as this would have direct implications for measured quantities. If the influence of a critical point is strong enough, hydrodynamic trajectories can be modified substantially, both in- and out-of-equilibrium. In general, what happens (see e.g., Ref.~\cite{Parotto:2018pwx}) is that the critical point attracts such trajectories, causing their clustering in its vicinity. 

In Fig.\ \ref{fig:schem_traj} we show a schematic comparison between trajectories with a weak or no critical point effect (left) and others where the effect is pronounced (right).  In the latter case, the trajectories accumulate around the critical region. 
In ideal hydrodynamics simulations, one would anticipate that the system is then much more likely to pass through the critical region when this effect is stronger. In this work, we will connect the strength of this effect to the size and shape of the critical region, and later explore to what extent it survives in the case of out-of-equilibrium viscous hydrodynamic simulations with various initial conditions. 

One can try and quantify how much the hydrodynamic trajectories are deformed by the presence of the critical point, by deriving $s/n$ with respect to $T$ or $\mu_B$. The total derivative of $s/n$ reads: 
\begin{align}\label{eq:dsdn}
    d(s/n) &= \frac{1}{n}\left( \dfrac{\partial s}{\partial T} dT + \frac{\partial s}{\partial \mu_B}d\mu_B \right) \\ \nonumber
    & \, \, \, \, - \frac{s}{n^2}\left(  \frac{\partial n}{\partial T}dT + \frac{\partial n}{\partial \mu_B} d\mu_B
    \right)  \, \, ,
\end{align}
from which one can easily see that
\begin{equation}\label{eq:muBseparation}
    \left.\left( \frac{\partial\mu_B}{\partial(s/n)}\right)\right|_T = \frac{n}{\left.\frac{\partial s}{\partial \mu_B}\right|_T - \frac{s}{n}\chi_2} \,\, ,
\end{equation}
and
\begin{equation}\label{eq:Tseparation}
    \left.\left( \frac{\partial T}{\partial (s/n)} \right)\right|_{\mu_B} = \frac{n}{ \frac{\partial s}{\partial T}|_{\mu_B} - s n\frac{\partial n}{\partial T}|_{\mu_B}} \, \, .
\end{equation}

Near the critical point along the crossover ($h=0$), the critical pressure can be written as
\beq\label{eq:criticalp}
P^{\textrm{crit}} = A r^{\beta\delta + \beta},
\eeq
where A is a constant, and $h\sim r^{\beta\delta}$ \cite{Schofield:1969zz}. The scaling of the pressure as $r\rightarrow 0$ can be used to estimate how each thermodynamic quantity behaves at the critical point (details in Appendix A). Both $s$ and $n$ scale with $r^{\beta}$, whereas the second-order derivatives diverge as $1/r^{\beta\delta-\beta}$. 
Substituting each leading term in $r$ into Eqs. (\ref{eq:muBseparation}) and (\ref{eq:Tseparation}) yields
\beq\label{eq:scalingofisentropeseparation}
\left.\left( \frac{\partial\mu_B}{\partial(s/n)}\right)\right|_T \sim \dfrac{1}{r^{\beta\delta}}, \quad
\left.\left( \frac{\partial T}{\partial (s/n)} \right)\right|_{\mu_B} \sim \dfrac{1}{r^{\beta\delta}},
\eeq

and we can conclude that the separation in $T$ and $\mu_B$ between isentropes goes to zero when the system exhibits criticality. Given the same set of initial conditions, there will be a larger density of trajectories in the critical circular region of Fig. \ref{fig:schem_traj} (right), when compared to the non-critical one (left). This is precisely the lensing effect we discuss in this work, which we have also observed in Ref.~\cite{Dore:2020jye}.

\section{Results: Equilibrium}

\subsection{Kurtosis and speed of sound} 
\begin{figure*}
    \centering
    \begin{tabular}{c c c}
    \includegraphics[width=0.33\linewidth]{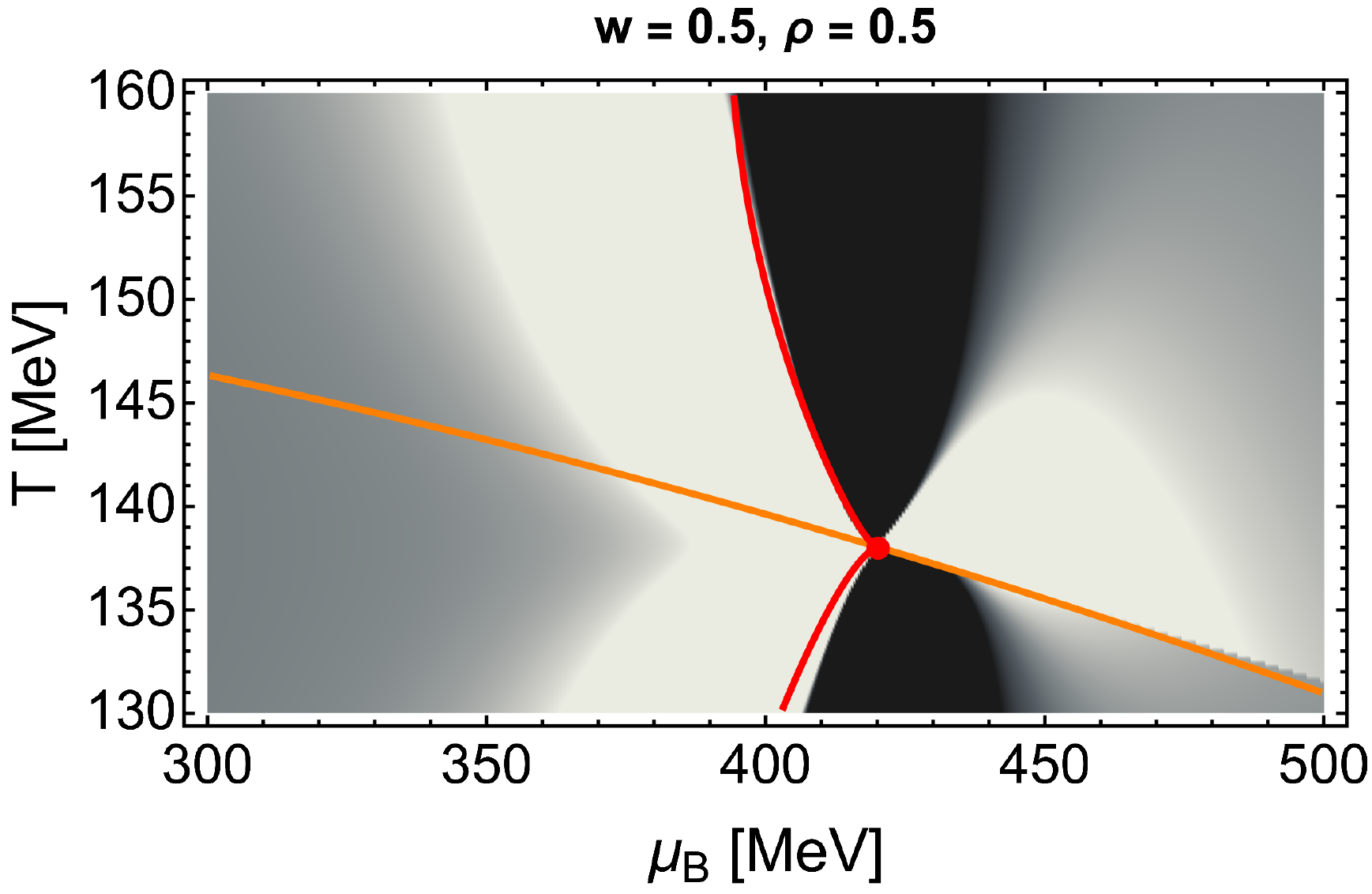} & \includegraphics[width=0.33\linewidth]{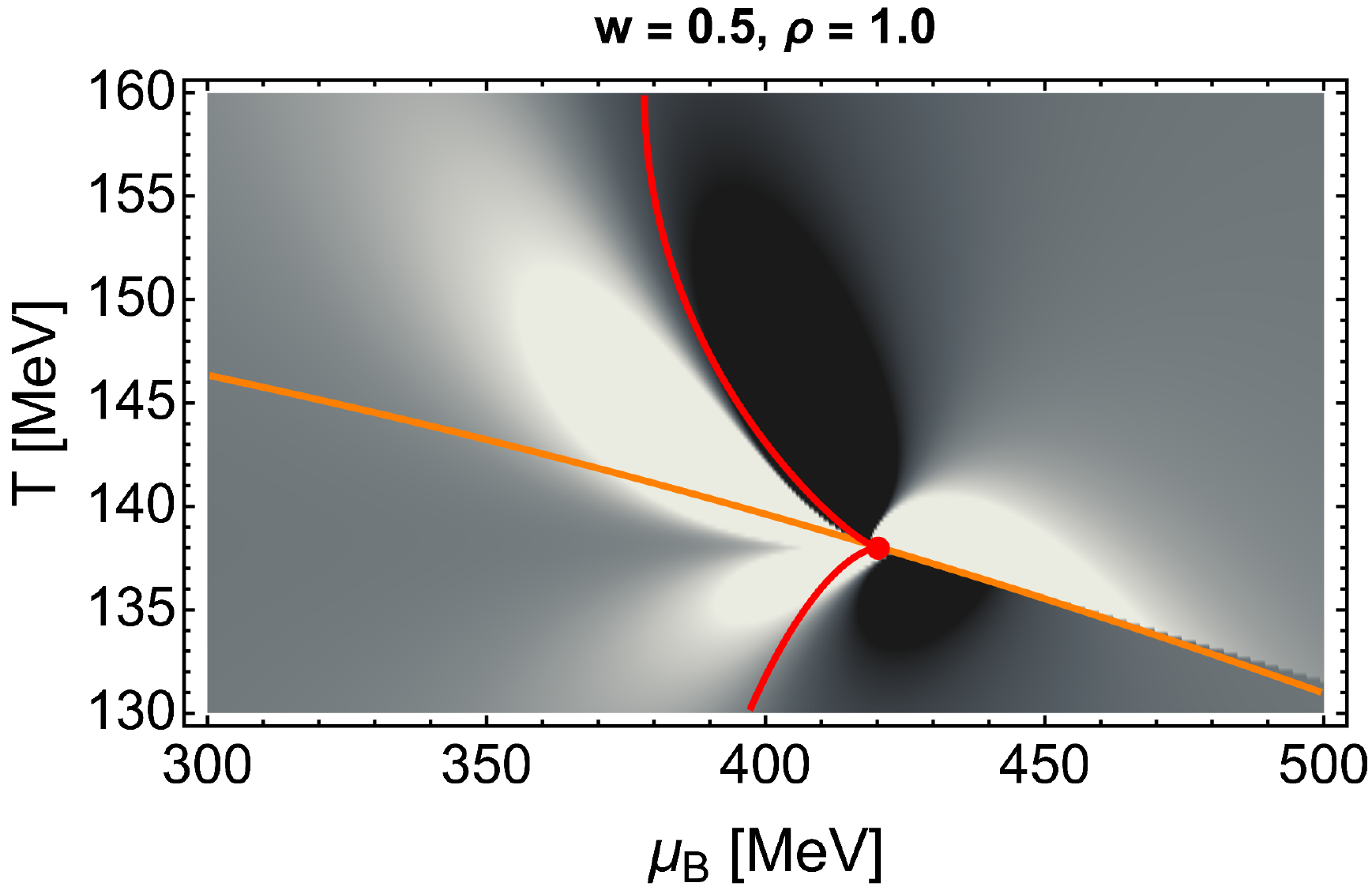} &  \includegraphics[width=0.33\linewidth]{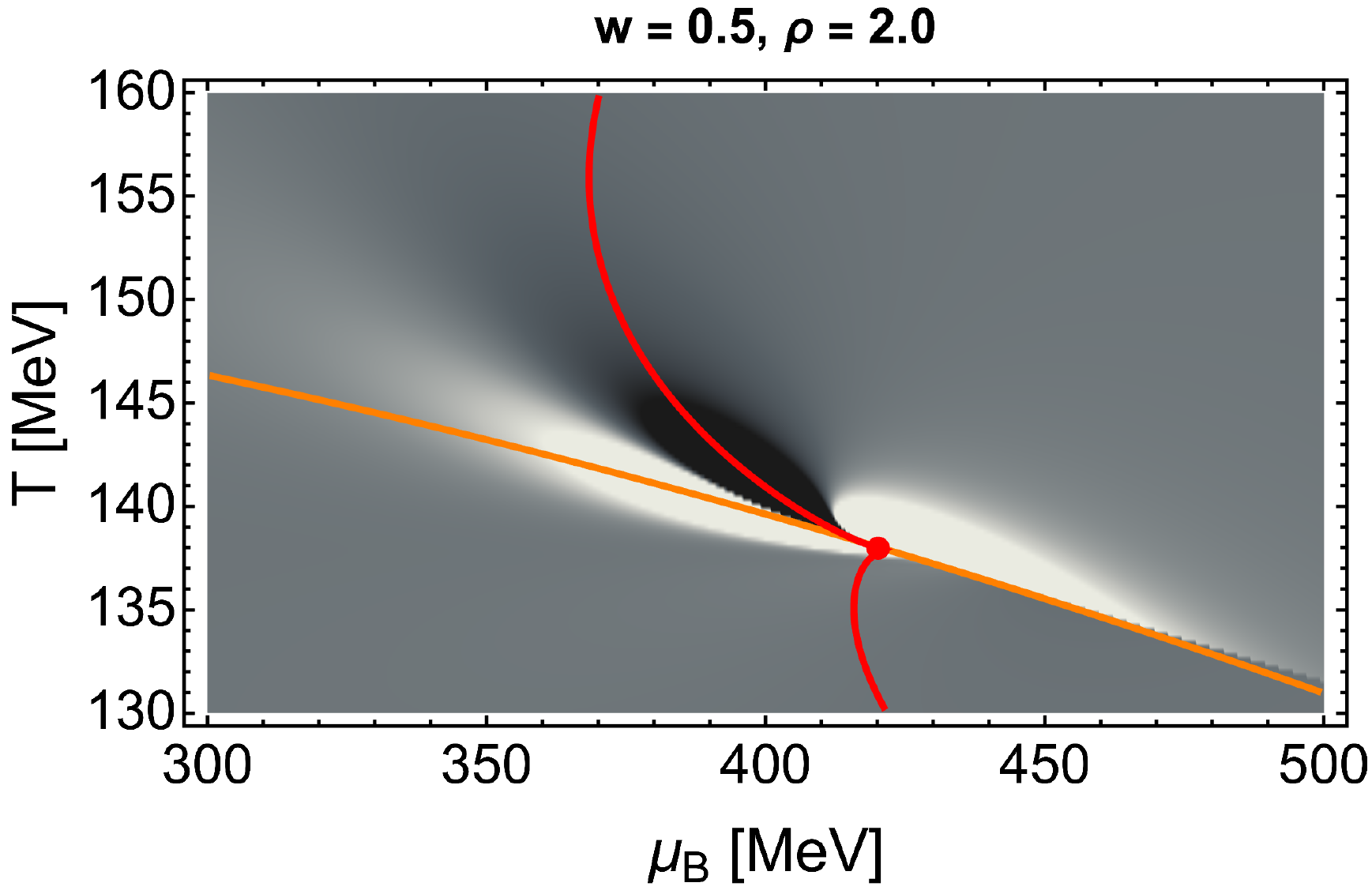}\\
    \includegraphics[width=0.33\linewidth]{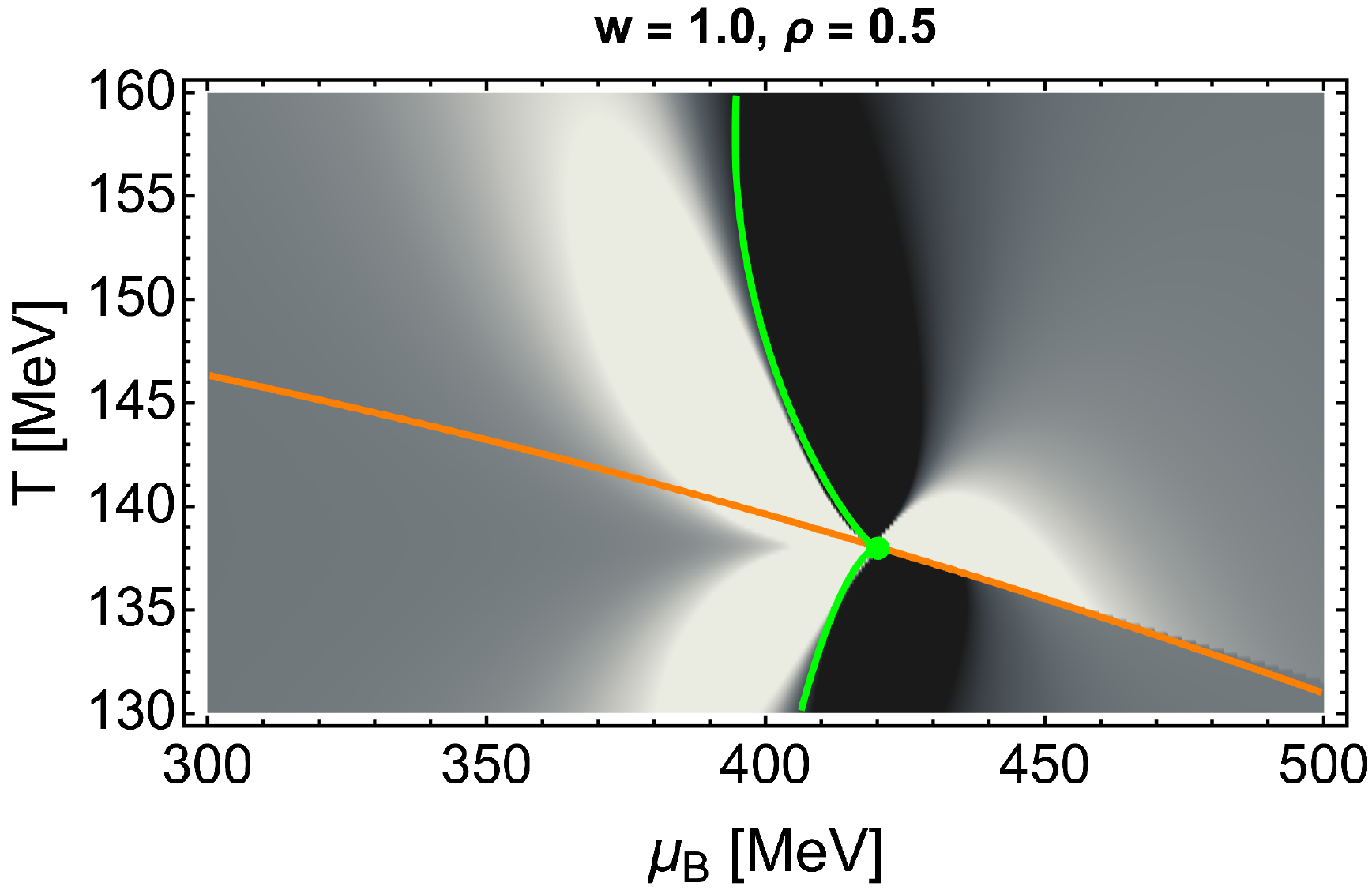} &
    \includegraphics[width=0.33\linewidth]{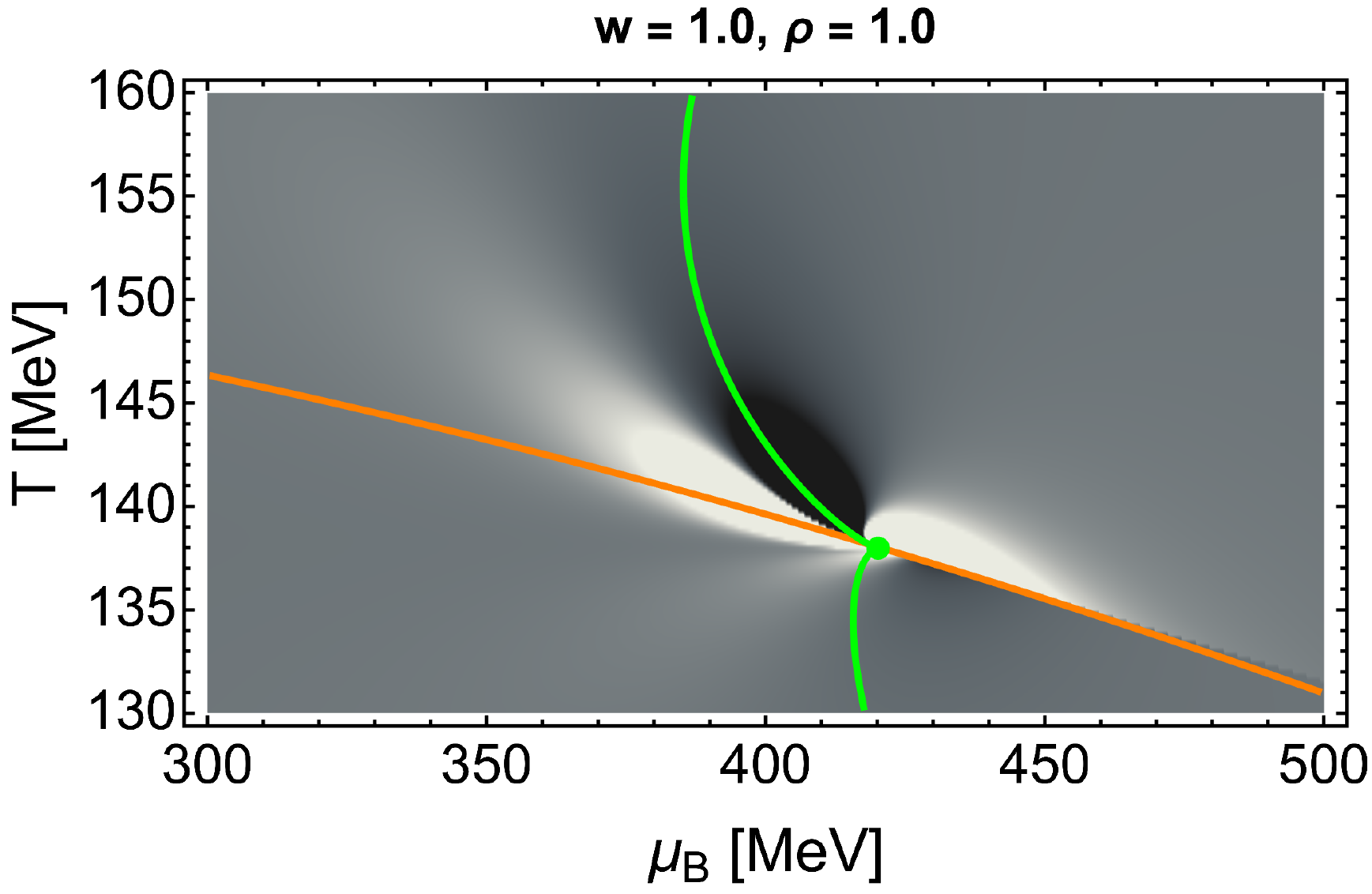} &  \includegraphics[width=0.33\linewidth]{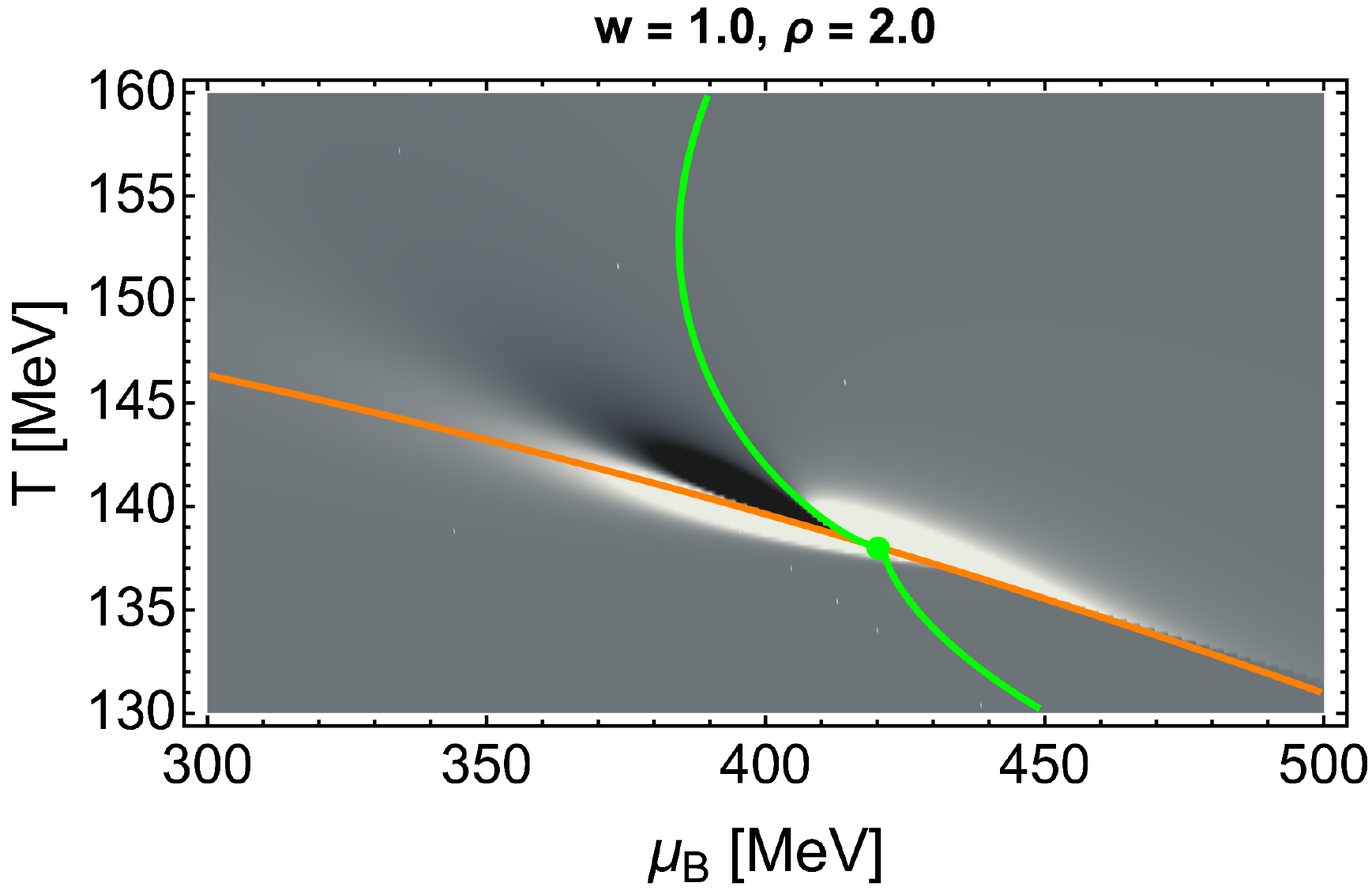}\\
    \includegraphics[width=0.33\linewidth]{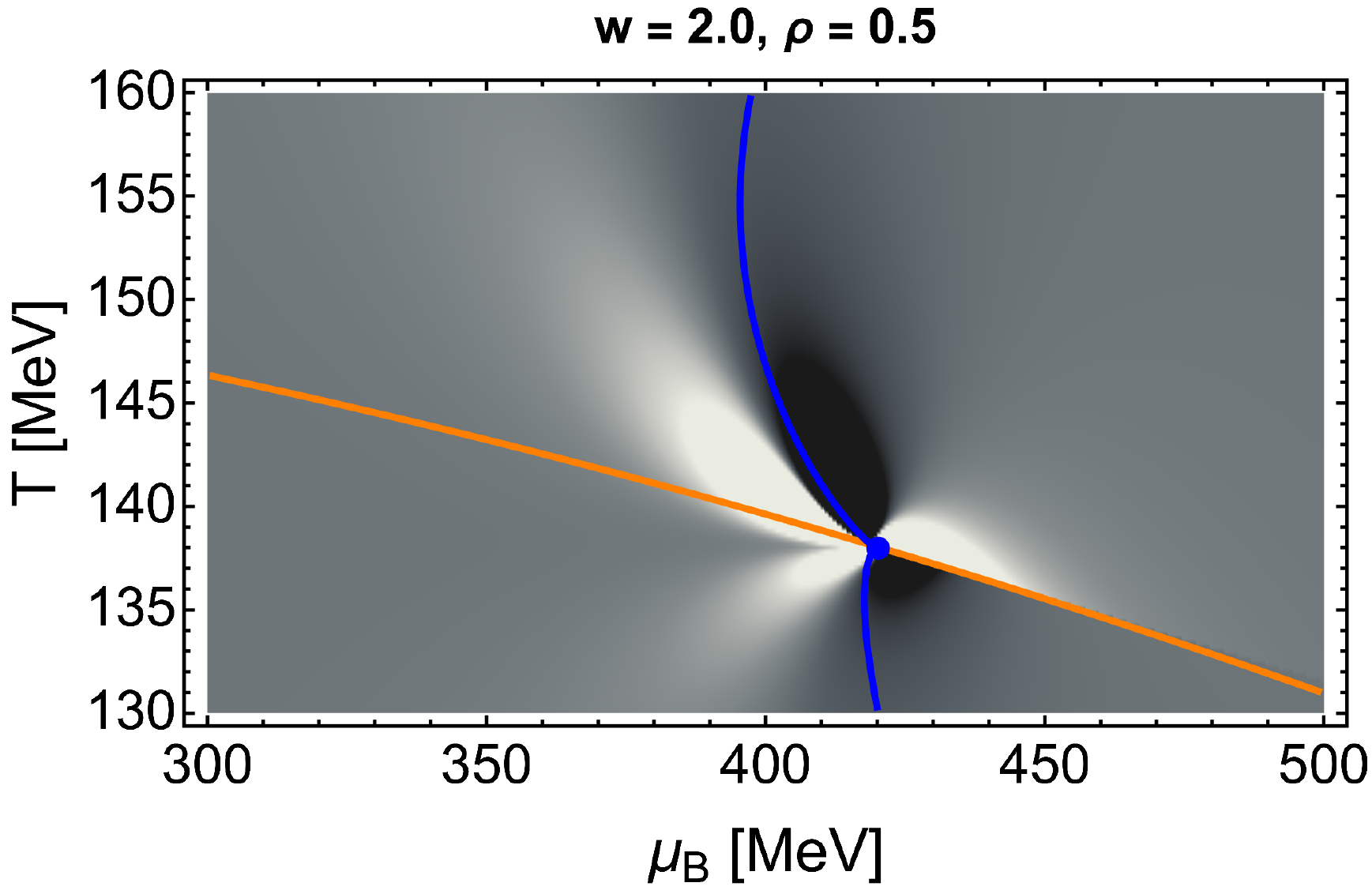} & 
    \includegraphics[width=0.33\linewidth]{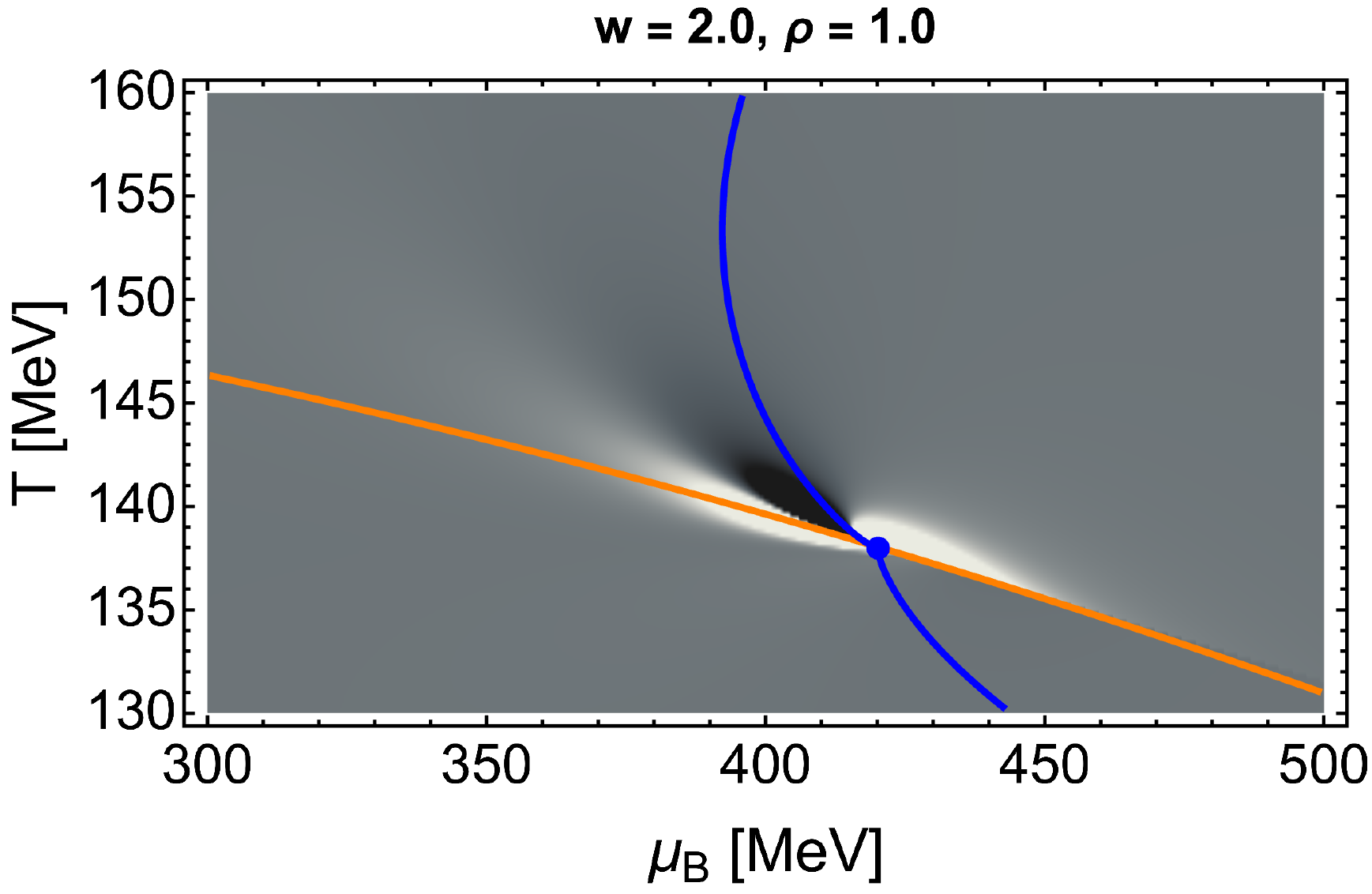} &  \includegraphics[width=0.33\linewidth]{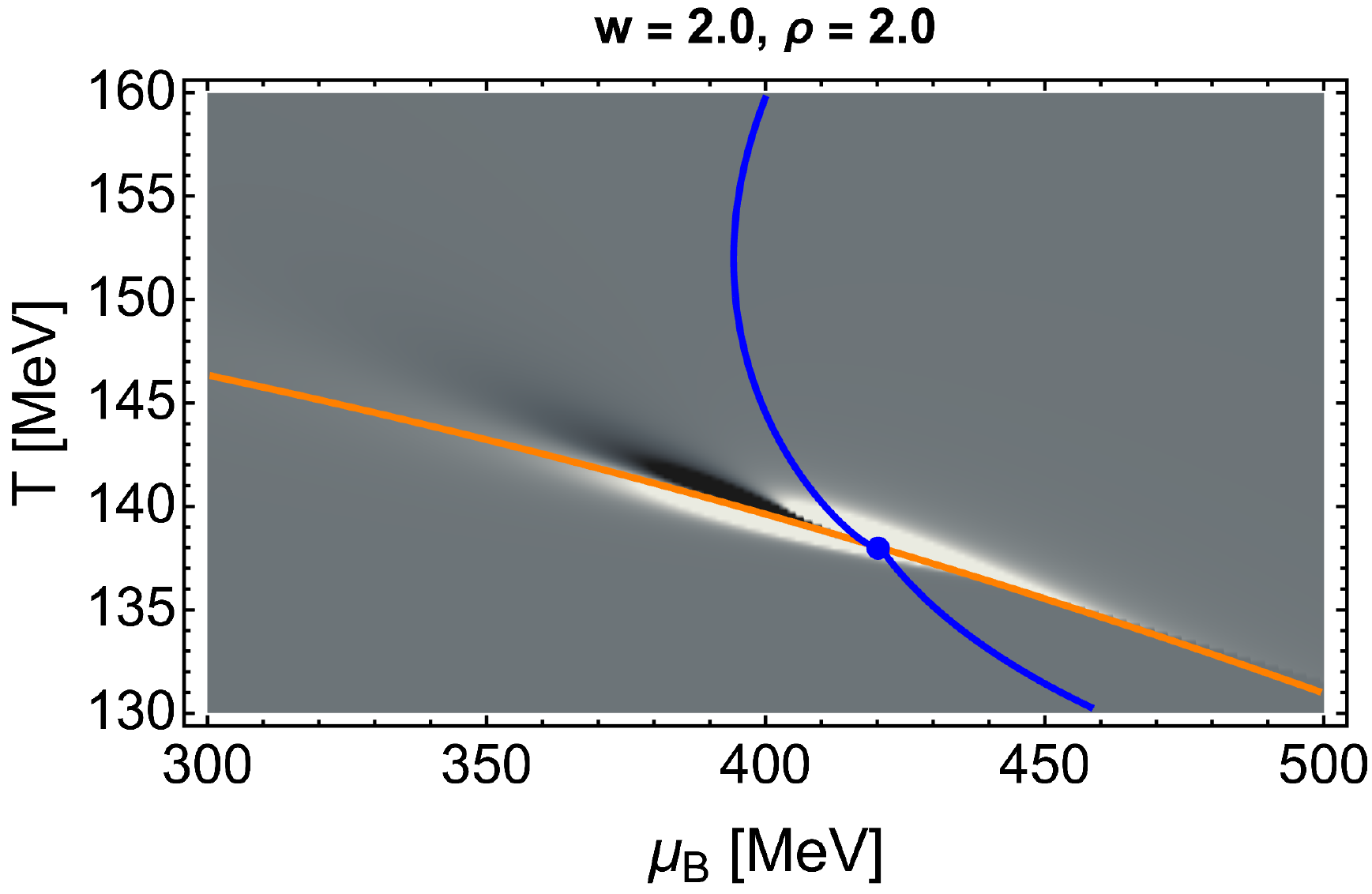}\\
    \includegraphics[width=0.33\linewidth]{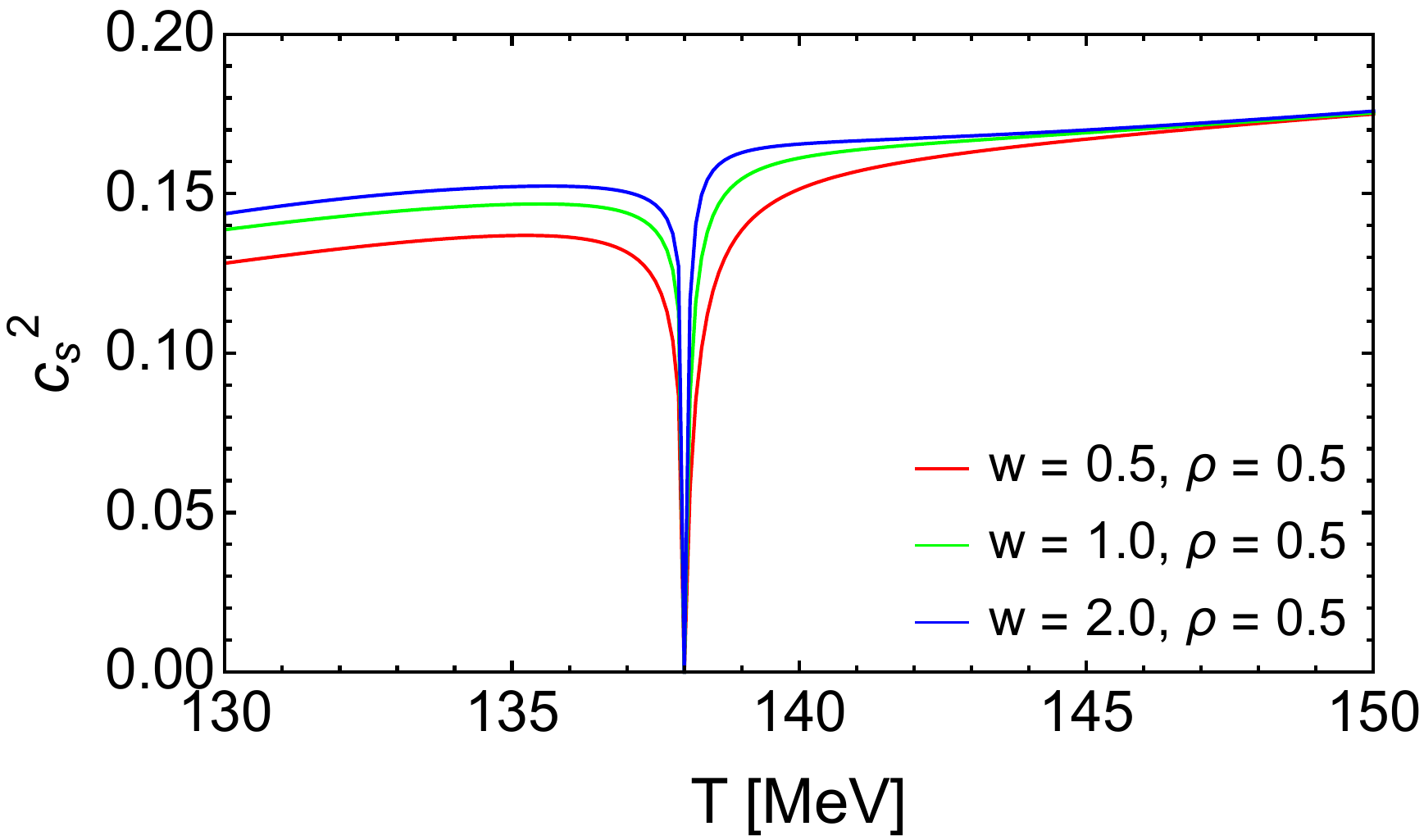} & 
    \includegraphics[width=0.33\linewidth]{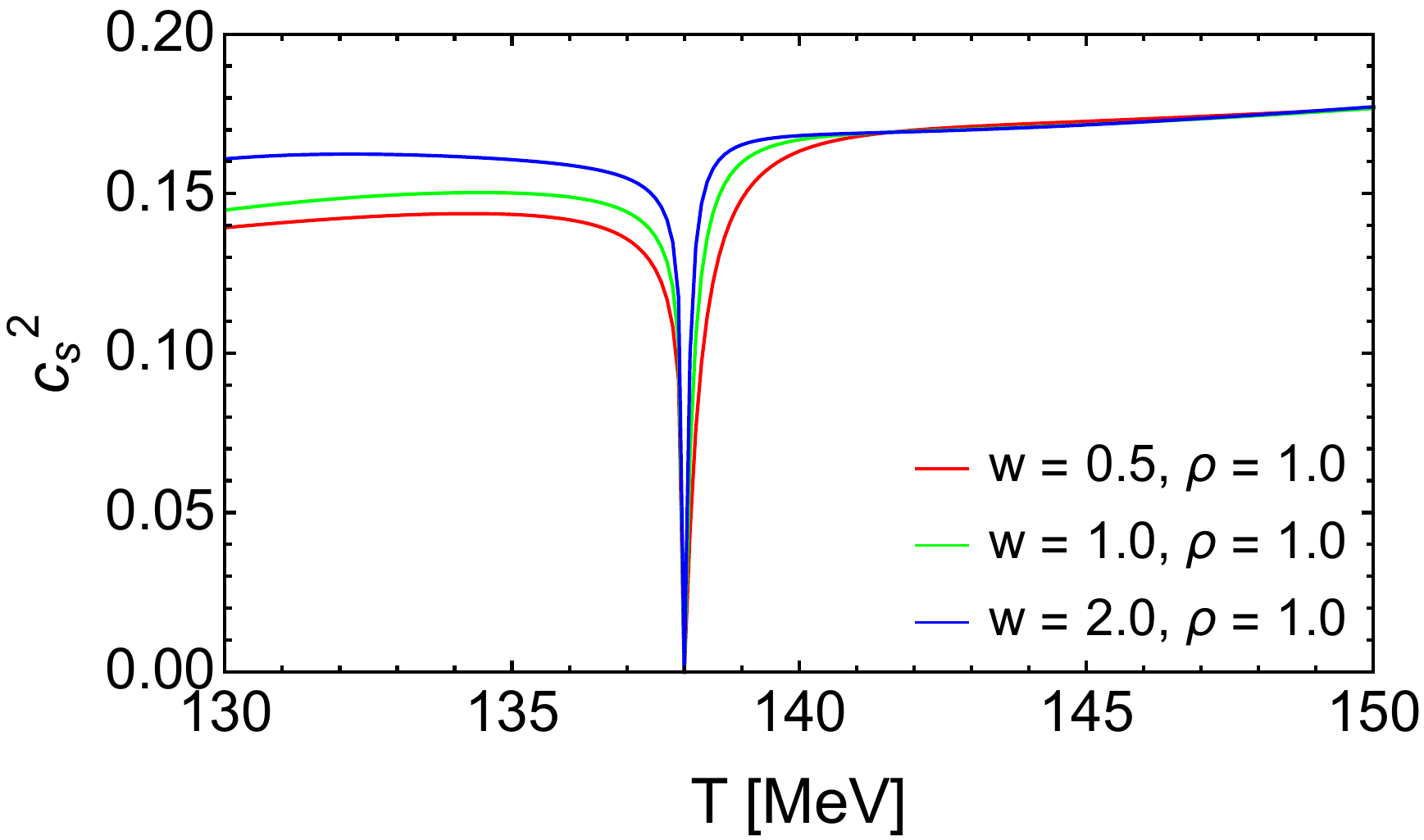}  & 
    \includegraphics[width=0.33\linewidth]{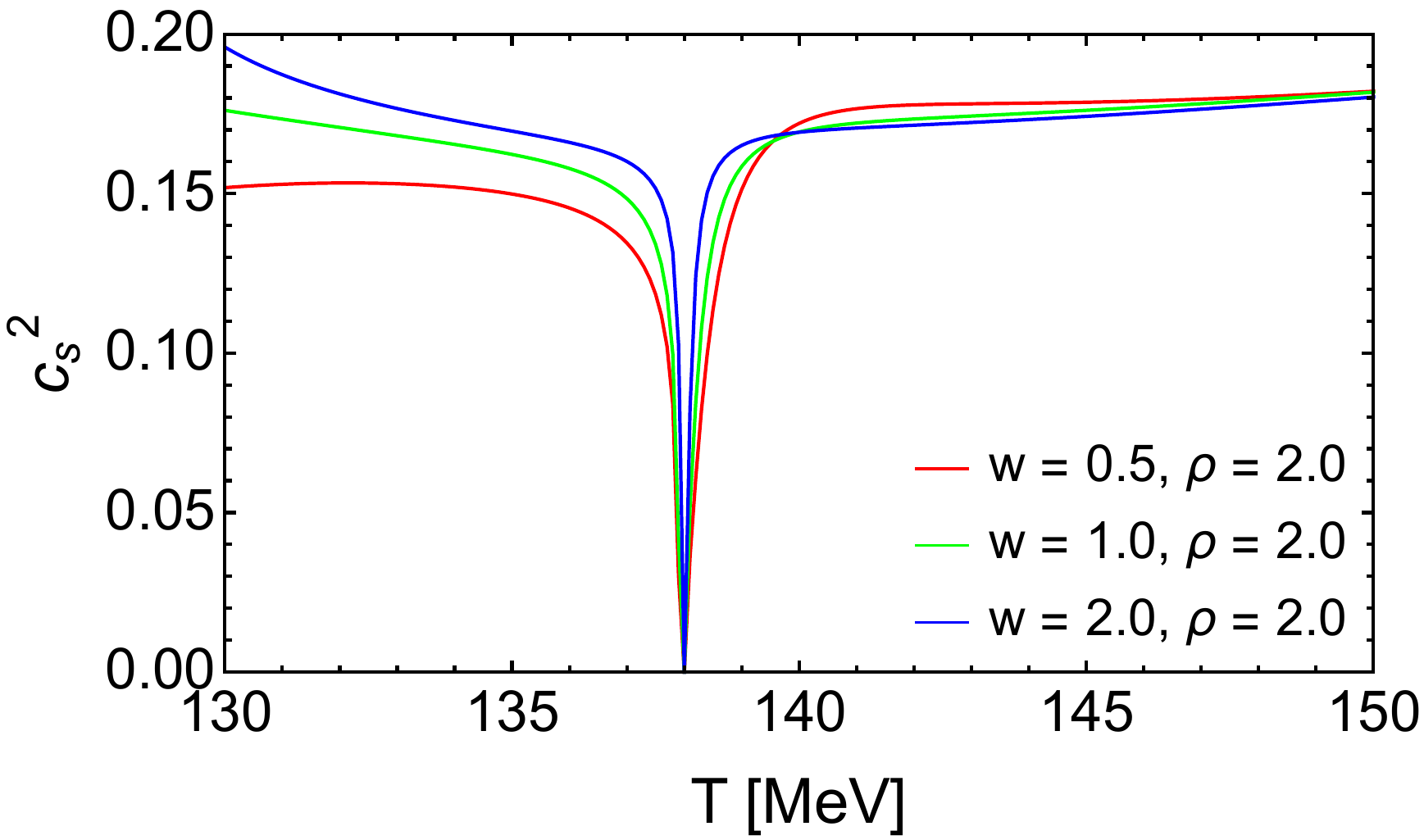}
    \end{tabular}
    \caption{Top three rows: contour plots of the fourth baryon number susceptibility $\chi_4^B$, with isentropic trajectories (solid red, green and blue lines) crossing the critical point. In all cases, we have $T_C=138$ MeV, $\mu_{BC}= 420$ MeV, $\alpha_1=4.6^\circ$, $\alpha_2-\alpha_1=90^{\circ}$. From top to bottom, left to right, we have $w= 0.5,1.0,2.0$, and $\rho=0.5,1.0,2.0$. The QCD transition line is represented by a solid orange line, while the critical point is represented by dots with the same color scheme as the isentropes. In the black regions $\chi_4^B<0$, in the white regions $\chi_4^B>0$, and in the gray regions $\chi_4^B \simeq 0$.
    Bottom row: speed of sound along the three isentropes in each column, with the same color scheme.}
    \label{fig:1eos}
\end{figure*}

In this Section, we will investigate how different realizations of the BEST Collaboration equation of state (i.e. different parameters in the Ising-to-QCD map) will influence the kurtosis and the critical lensing effect. Because of the numerous complications in studying the physics of heavy-ion collisions in the vicinity of the critical point, it is crucial to understand the interplay between the features of the equation of state in the critical region, the evolution trajectories of hydrodynamic simulations, and observables such as net-proton fluctuations. A particularly important role is played by the speed of sound, which is expected to vanish at the critical point. Although the scaling behavior of how $c_s^2\rightarrow 0$ is known, sub-leading contributions might have an important role, and thus modify the speed of sound over a sizeable portion of the system evolution. Relativistic hydrodynamics is quite sensitive to this change in $c_s^2$ when the trajectory goes through the critical region, due to the connection between $c_s^2$ and $\zeta T/w$~\cite{Dore:2020jye}. 

In Fig.\ \ref{fig:1eos} we show the kurtosis across the $\left\{T,\mu_B\right\}$ plane for different parameters sets of the EoS. We fix in all cases:
\begin{itemize}
\item $T_C = 138$ MeV
\item $\mu_{BC} = 420$ MeV
\item $\alpha_1 = 4.6^{\circ}$
\item $\alpha_2 = 94.6^{\circ}$
\end{itemize}
and consider all possible combinations of $w = 0.5, 1.0, 2.0$ and $\rho = 0.5, 1.0, 2.0$. The same parameters were studied in Ref.~\cite{Mroczek:2020rpm} (Fig.1), and were chosen to produce varying critical regions that extend across the transition line (i.e. across the $\mu_B$ direction of the phase diagram), perpendicular to the transition line (i.e. across the $T$ direction of the phase diagram), or a combination of both. Following Ref.~\cite{Mroczek:2020rpm}, we use the magnitude of $\chi_4^B$ to categorize the size and shape of the critical region. The region where the critical contribution to $\chi_4^B$ is sizeable, i.e. the critical region, is shown in white (large and positive) or black (large and negative). The gray regions indicate a negligible critical contribution, where the effect of the critical point is absent. Some of us in Ref.~\cite{Mroczek:2020rpm} described the connection between the size of the critical region and the parameters $w$ and $\rho$. We found its extent in the temperature direction, at constant $\mu_B$ to be $\Delta^\textrm{crit}_T \sim w^{-3/7}$, and in the chemical potential direction, along the transition line $\Delta^\textrm{crit}_{\mu_B} \sim \rho w^{1/7}$. 

In  Fig.~\ref{fig:1eos}, we also show the isentropic trajectory that passes through the critical point, in either red, green, or blue. Isentropic trajectories are characterized by having a constant entropy-to-baryon-number ratio $s/n$, which is a conserved quantity in ideal hydrodynamics. 
If a collision could be well-described without viscous hydrodynamics, then the initial condition would only be a point in the $\left\{T,\mu_B\right\}$ plane, after which the system would expand and cool along the specific isentropic trajectory defined by the initial condition.

Near the critical point along the crossover, we can use the scaling behavior of the critical contribution to the pressure $P^\textrm{crit} \sim r^{\beta\delta + \beta}$ and the map between Ising and QCD variables to determine that the separation between isentropes along the $\mu_B$ direction scales with the EoS parameters $w$ and $\rho$ as (detailed derivation shown in Appendix B):
\beq
\dfrac{d(s/n)}{d\mu_B} \sim \frac{1}{ (w \rho)  r},
\eeq
Since $d(s/n)/d\mu_B$ diverges with $1/r $ as $r\rightarrow 0$ with an overall factor directly proportional to $(w \rho)^{-1},$ we expect EoS generated from smaller $w$ and $\rho$ values to display a more dramatic lensing effect.  

\begin{figure}
    \centering
    \includegraphics[width=.95\linewidth]{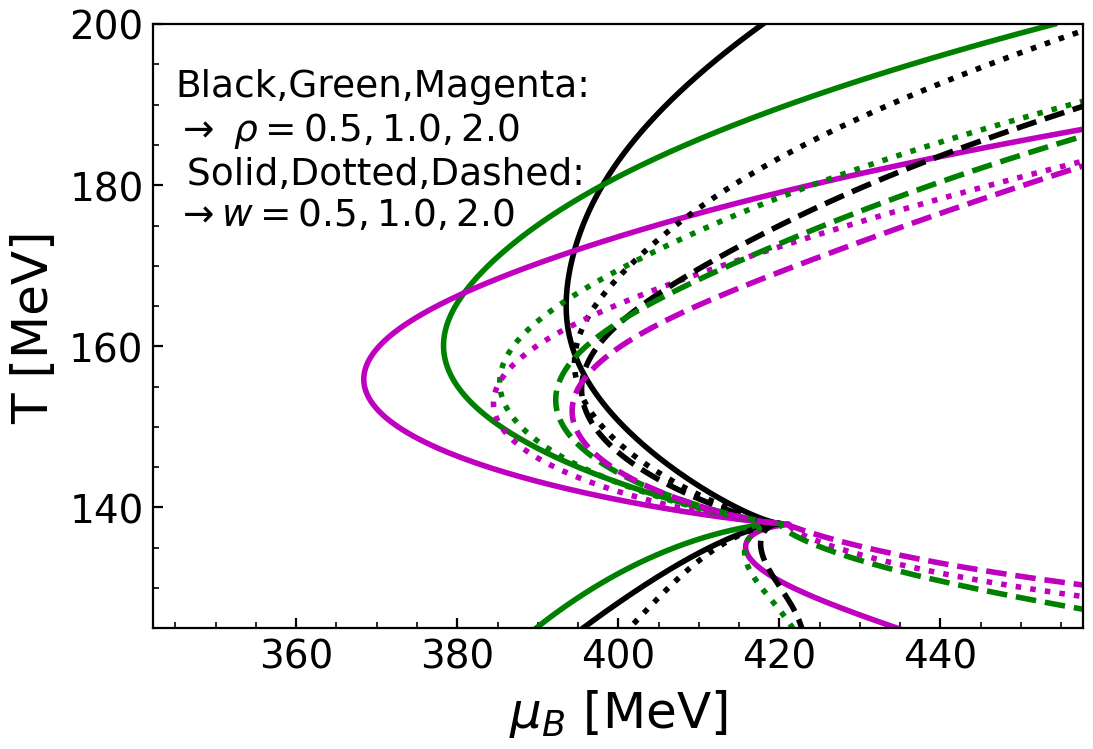}
    \caption{The isentropes from all EoS's in Fig.\ \ref{fig:1eos}, on a single plot. The solid, dotted, dashed lines correspond to $w = 0.5, 1.0, 2.0$ respectively. The black, green, magenta lines correspond to $\rho = 0.5,1.0,2.0$ respectively.}
    \label{fig:allIsen}
\end{figure}

By comparing our choice of parameters with the shape of the isentropic trajectory, we can confirm our predictions for the strength of the lensing effect. We find that critical regions that extend farther along the $T$ direction (corresponding to  smaller $w$ and $\rho$) generally have a more pronounced kink near the critical point.  This is not the case when the critical region extends mostly along the $\mu_B$ direction (larger $w$ and $\rho$). We plot all these isentropic trajectories together in Fig.~\ref{fig:allIsen}, where the effect is made even more evident. This shows that the critical lensing effect is not only affected by the size of the critical region, but also by its shape. This is because, in ideal hydrodynamics, it is the speed of sound that determines the evolution of the system.

In the bottom row of Fig.\ \ref{fig:1eos}, we show the speed of sound $c_s^2$ along the different isentropes for fixed $\rho$, while varying $w$. In all cases, it is apparent that larger values of $w$ lead to narrower dips in $c_s^2$. The value of $\rho$ seems to affect the low-$T$ region (below the critical point) only. This is in line with the fact that, as observed, the extent of the critical region in the temperature direction is $\Delta^\textrm{crit}_T \sim w^{-3/7}$, thus independent of $\rho$.

\begin{figure*}
    \centering
    \begin{tabular}{c c c}
    \includegraphics[width=0.33\linewidth]{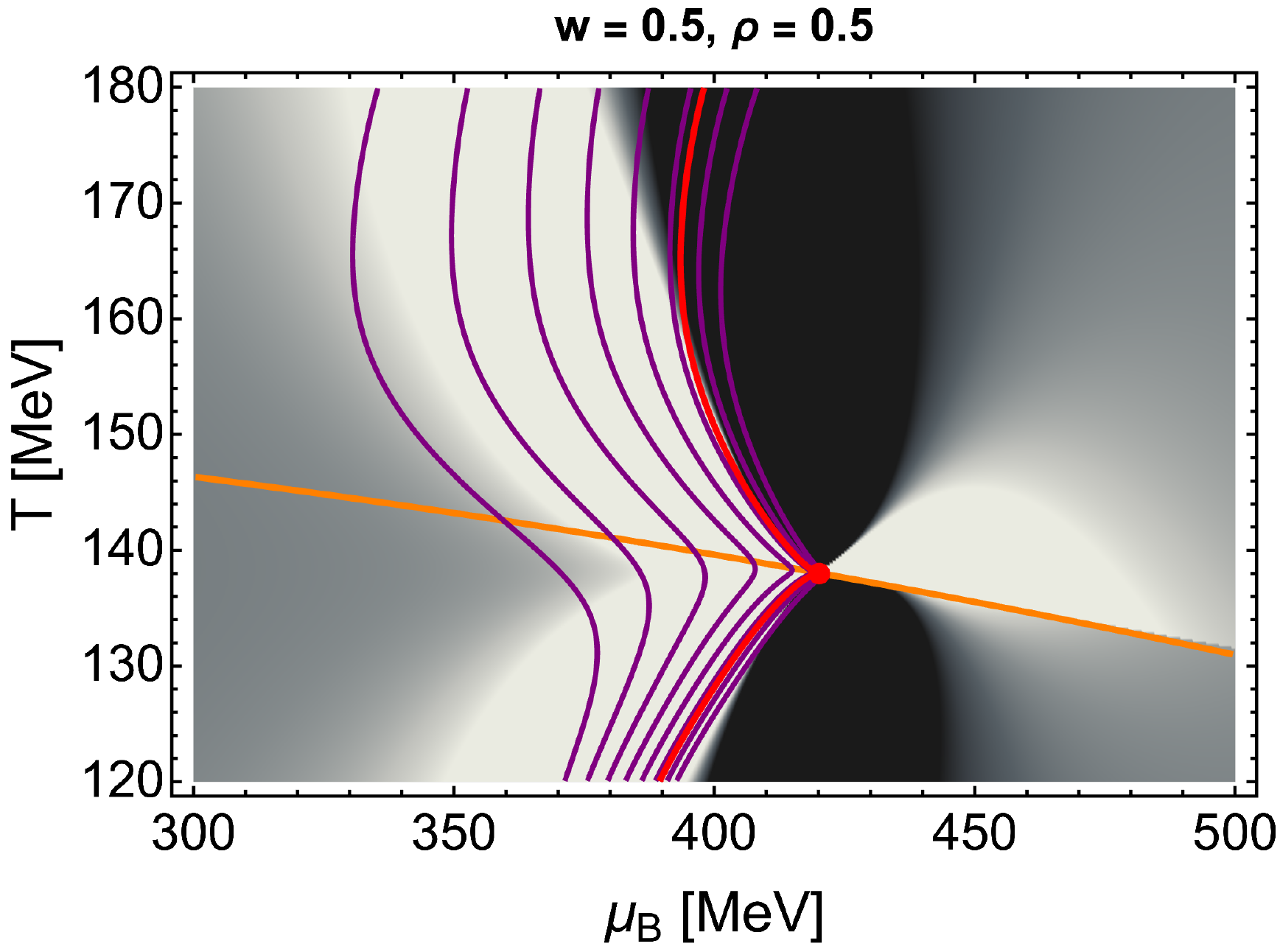} & \includegraphics[width=0.33\linewidth]{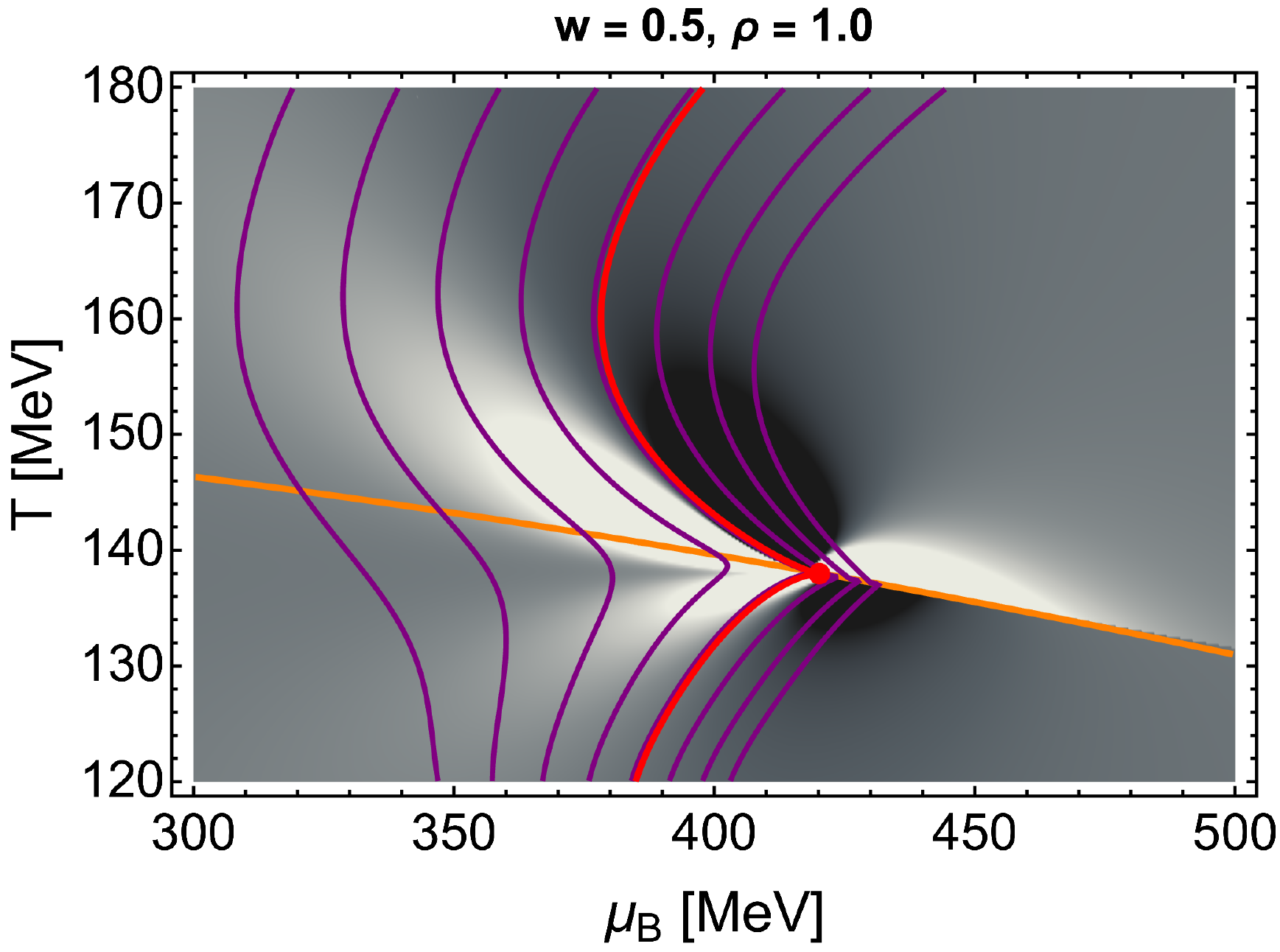} &  \includegraphics[width=0.33\linewidth]{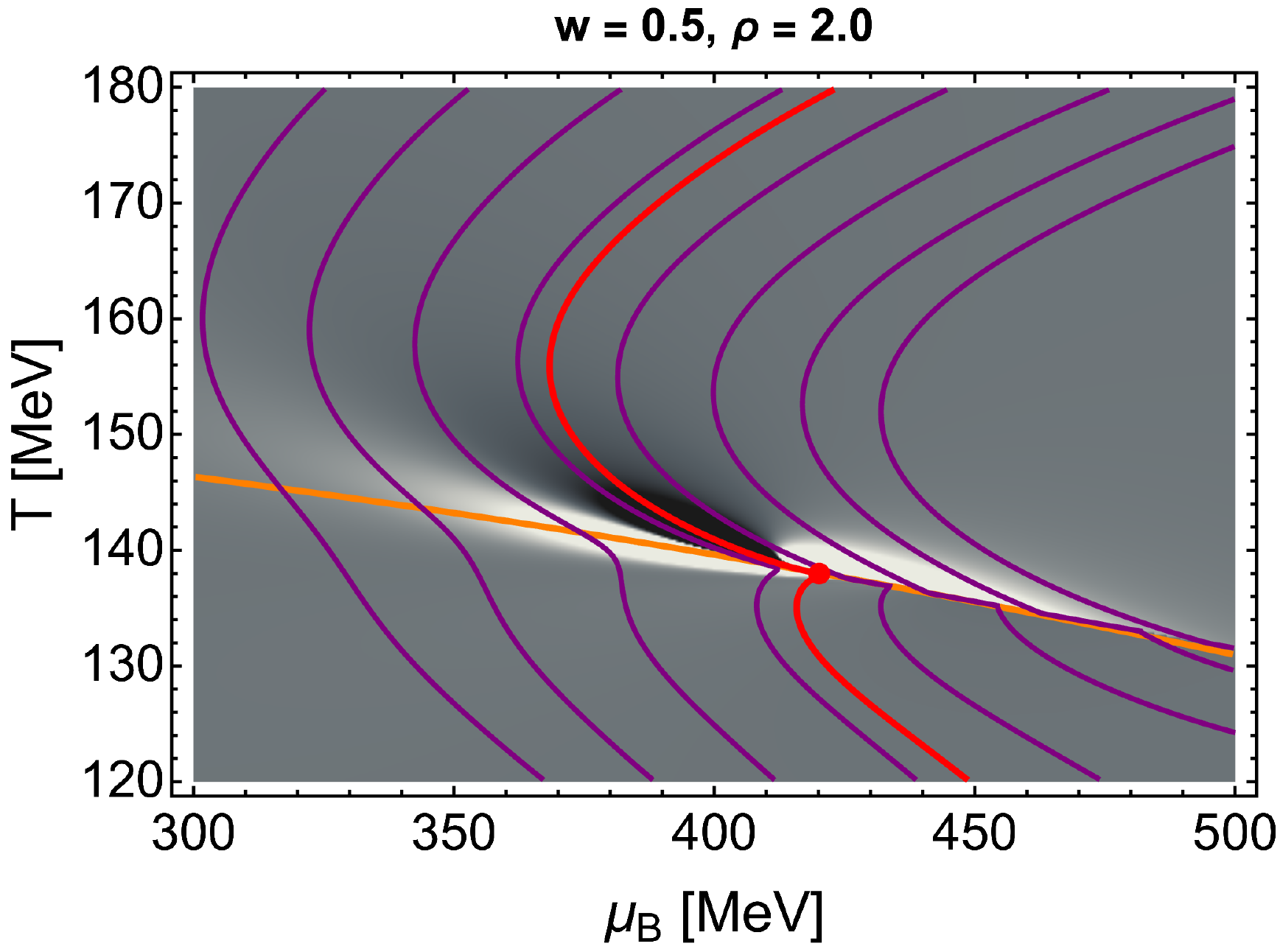}\\
    \includegraphics[width=0.33\linewidth]{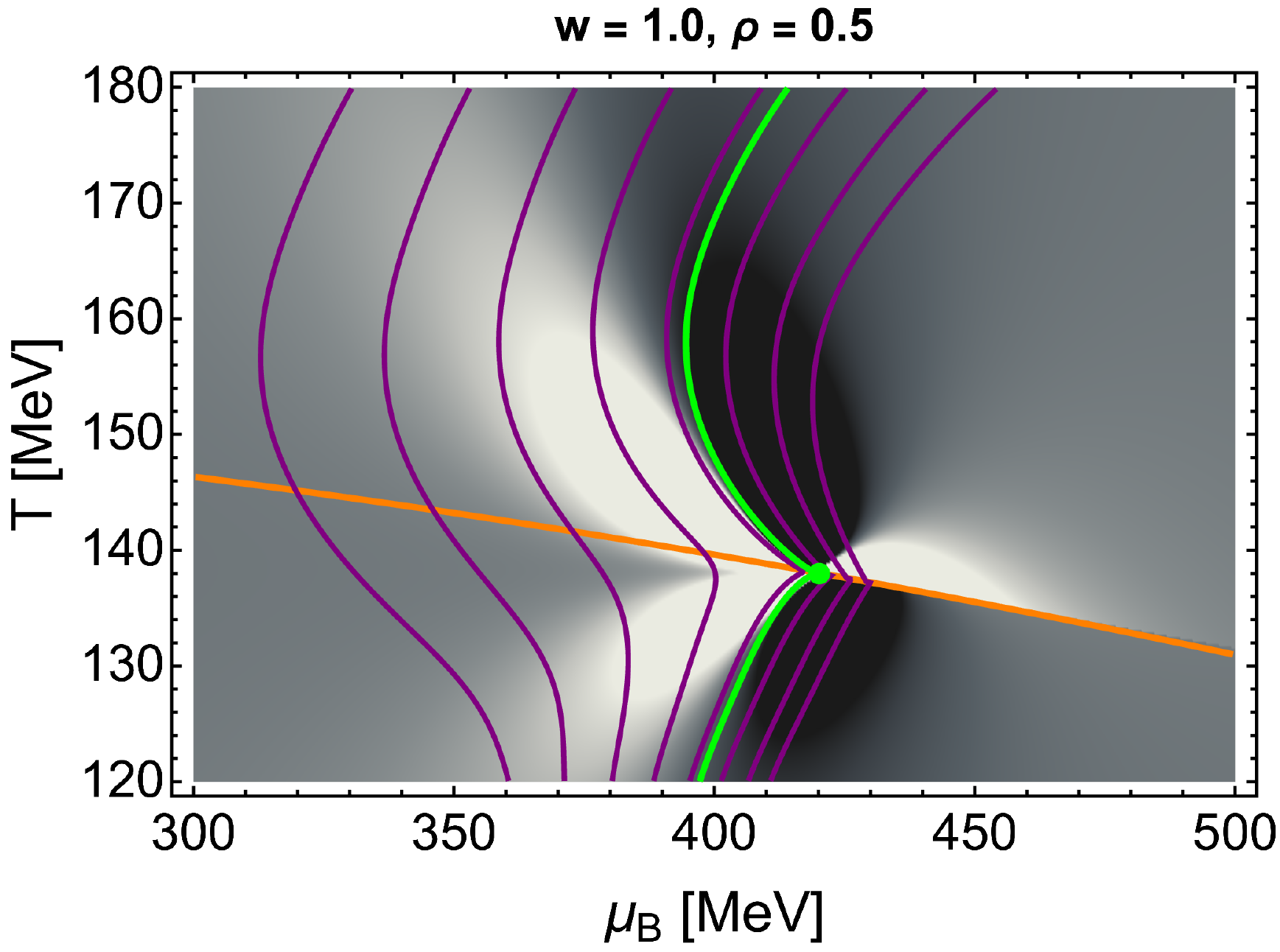} &
    \includegraphics[width=0.33\linewidth]{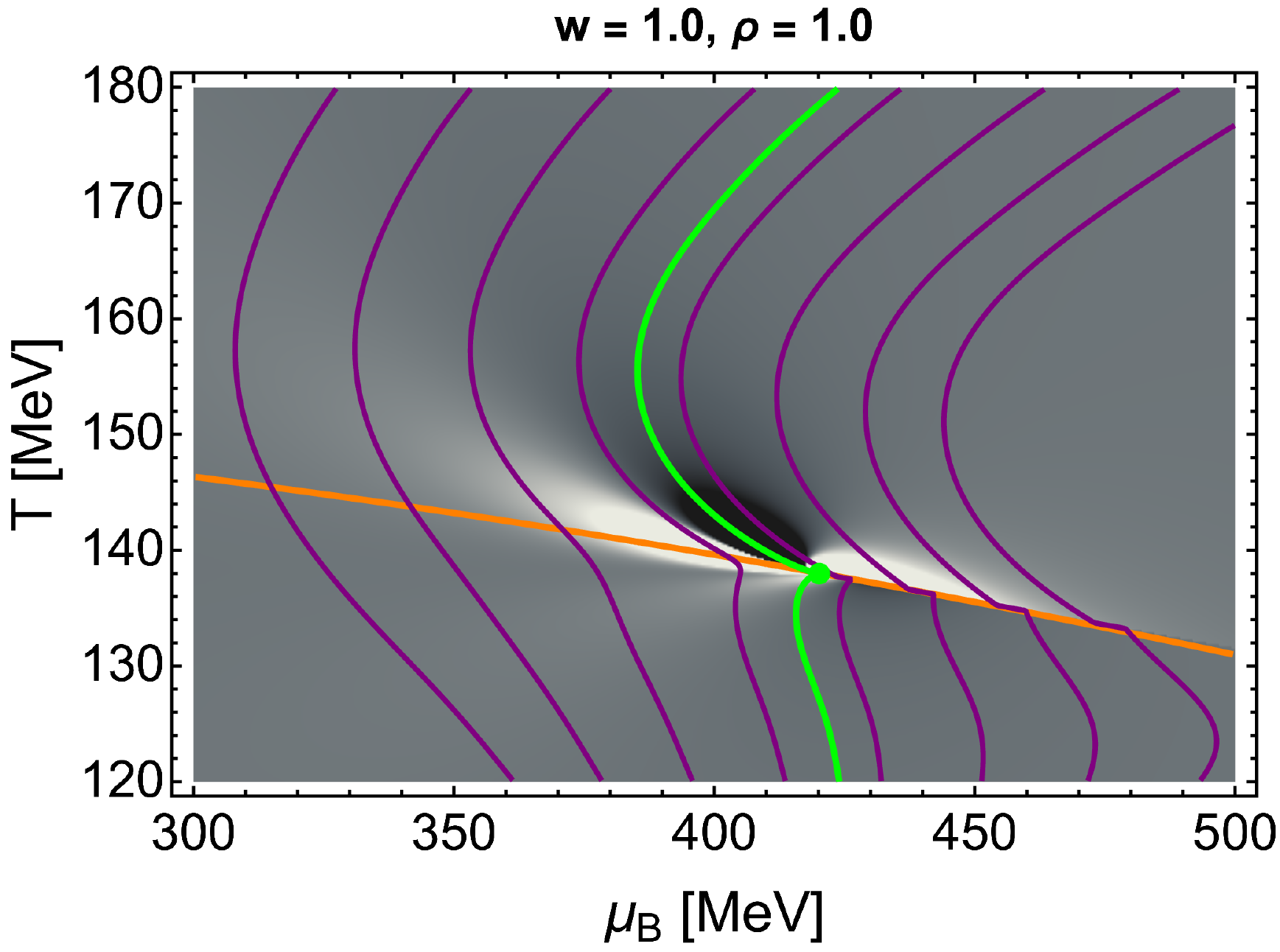} &  \includegraphics[width=0.33\linewidth]{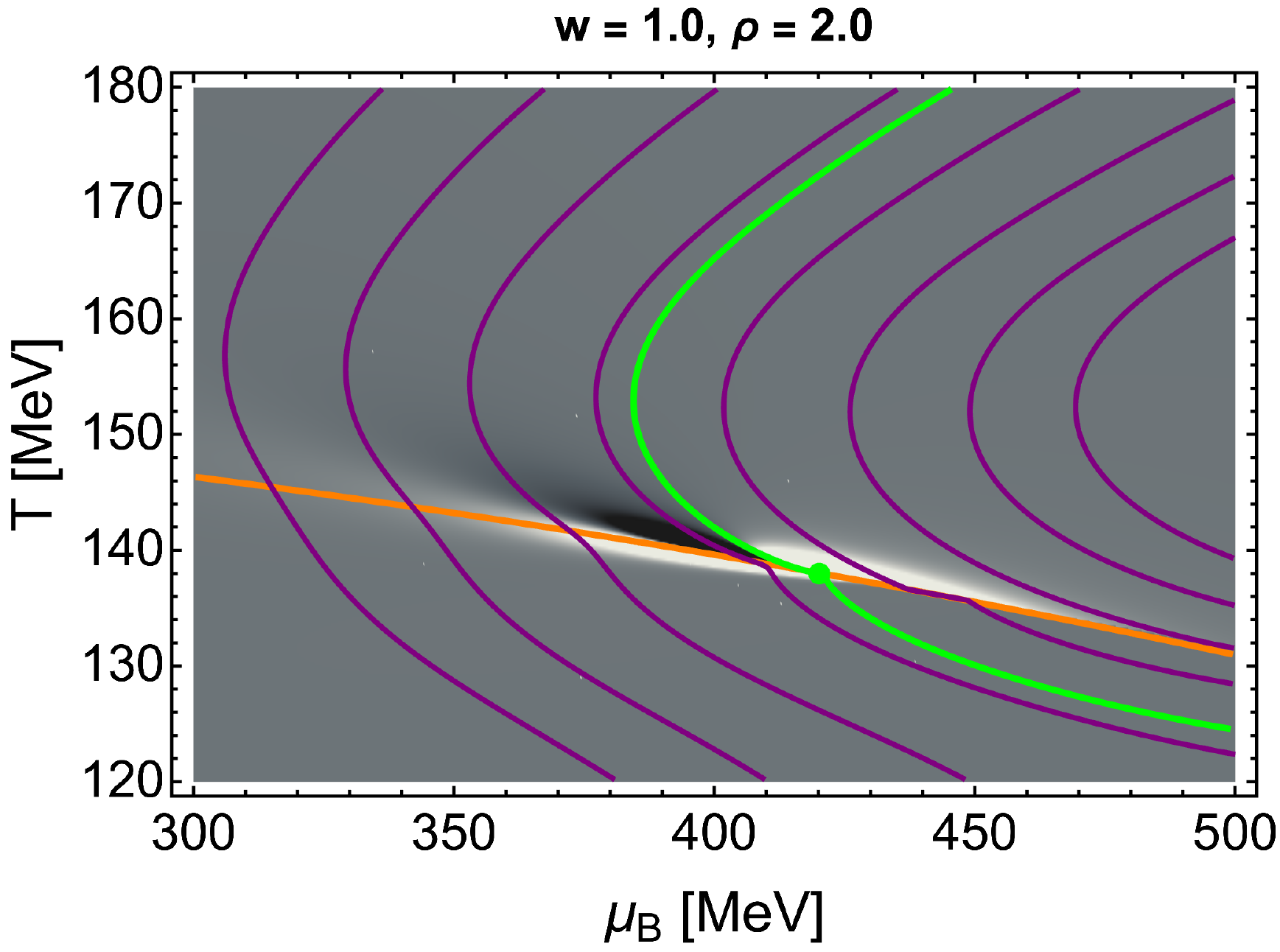}\\
    \includegraphics[width=0.33\linewidth]{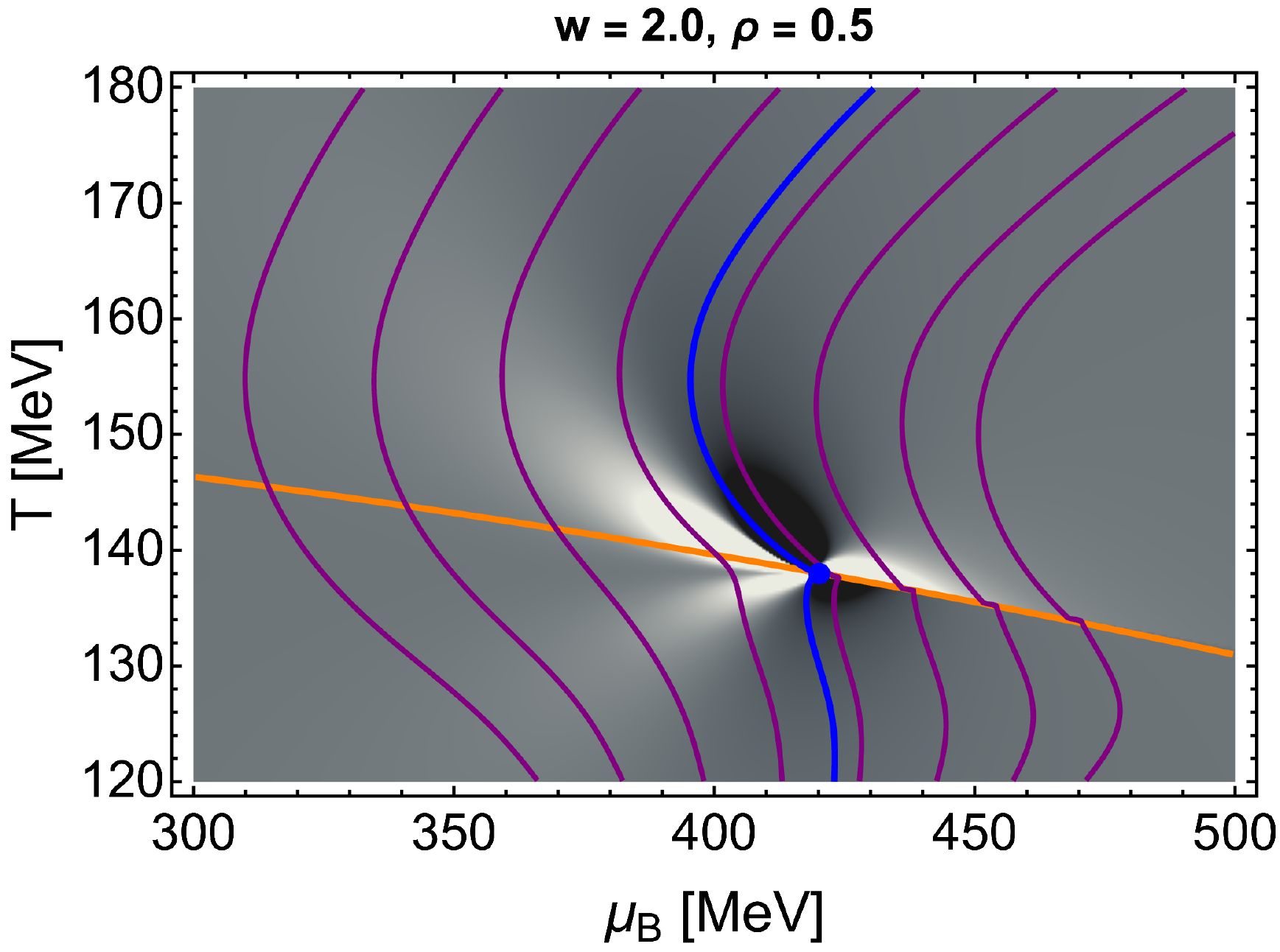} & 
    \includegraphics[width=0.33\linewidth]{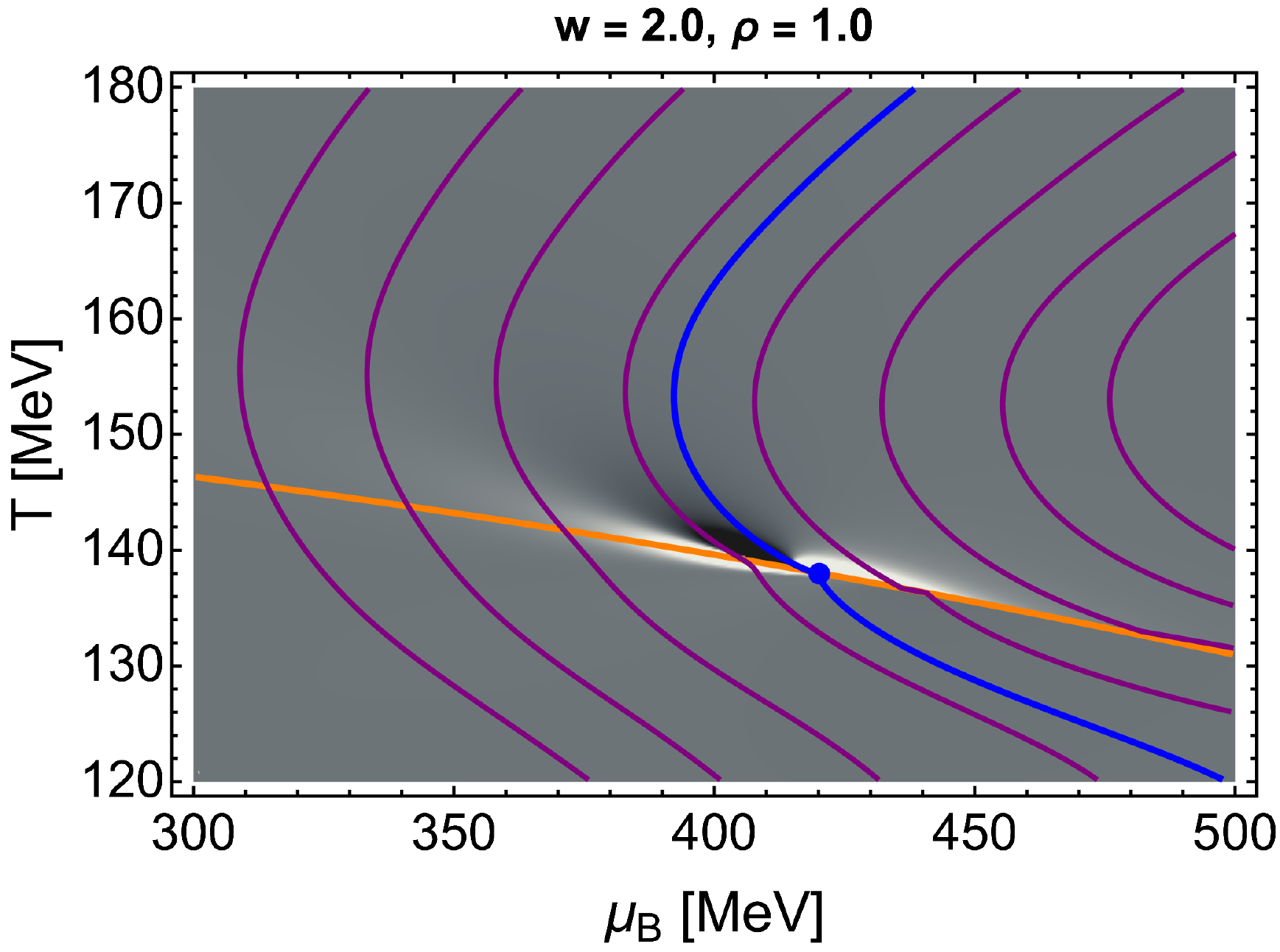} &  \includegraphics[width=0.33\linewidth]{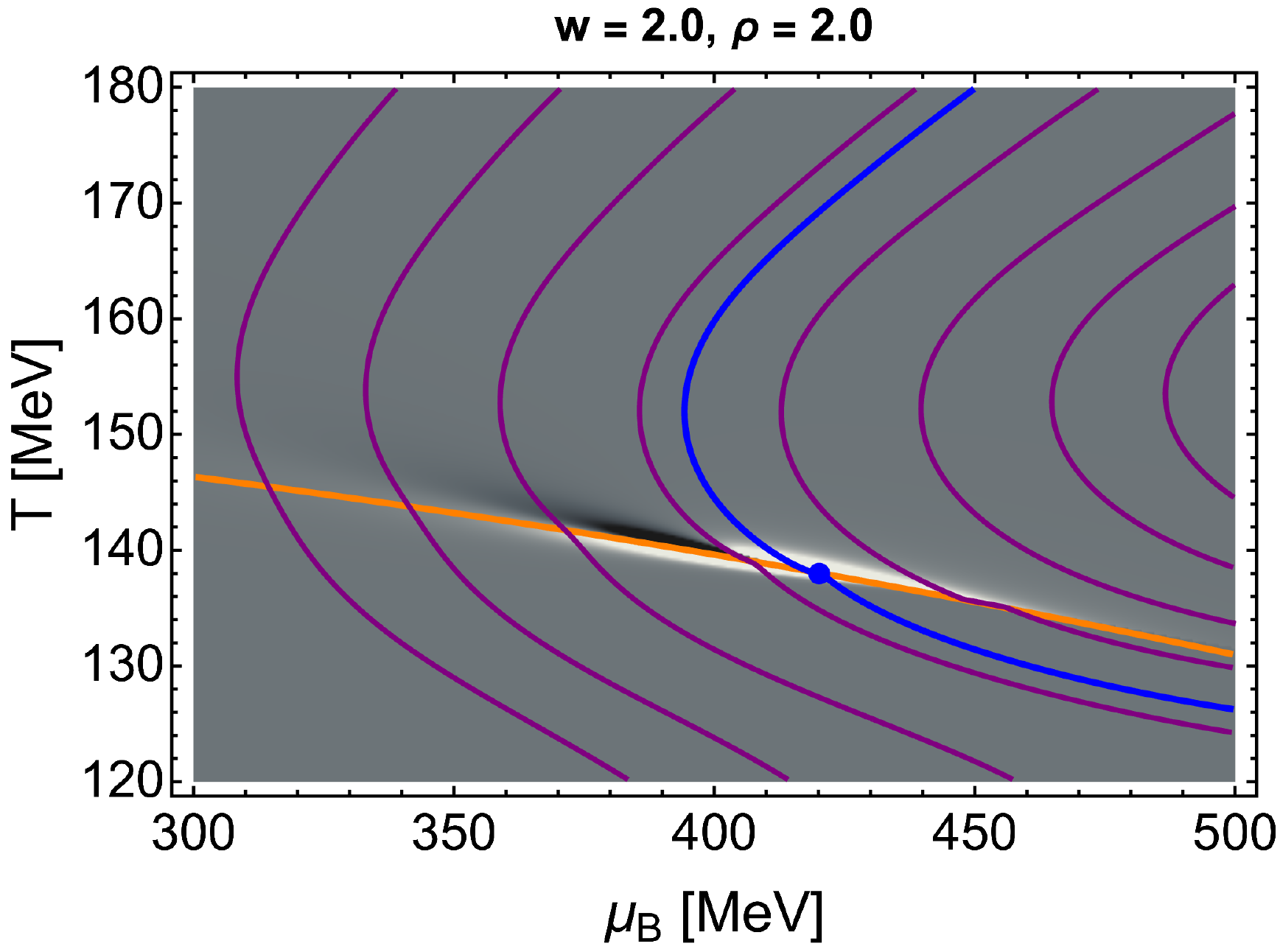}
    \end{tabular}
    \caption{Same as the top three rows of Fig.\ref{fig:1eos}, with additional isentropes corresponding to $s/n_B=(23.1,21.2,19.5,18.0,16.7,15.6,14.7,14.0)$ (purple solid lines). The isentropes crossing the critical point are shown in the same color scheme as in Fig.\ \ref{fig:1eos}.}
    \label{fig:critical_lensing}
\end{figure*}

To better visualize the critical lensing effect, in Fig.\ \ref{fig:critical_lensing} we plot different isentropic trajectories at fixed intervals of $s/n_B$, for the same parameter choices we previously showed in Fig. \ref{fig:1eos}.

A picture consistent with that of Fig.\ \ref{fig:1eos} emerges, in which the cases where the trajectories are more deformed coincide with those where they are also more evidently amassed, i.e. smaller values of $w$ and $\rho$.

An interesting point that can be seen from the figures is that the critical lensing effect exists both on the the first-order \textit{and} crossover sides of the critical point. We will see in Sec.\ \ref{sec:hydroTraj} that this is still the case when including out-of-equilibrium effects. Though other works have investigated critical lensing in equilibrium, they mostly have done so in the case of first-order phase transitions \cite{Stephanov:1998dy,Stephanov:2004wx,Nonaka:2004pg,Asakawa:2008ti}. In Ref.~\cite{Nonaka:2004pg}, the lensing effect is shown for trajectories that start both on the crossover side and first-order side, but in the end almost always pass through either the critical point or first-order side of the transition region (see e.g., Fig. 4 in that paper). The authors of Ref.~\cite{Asakawa:2008ti} study the effects of turning on or off 
the critical point, while using a single realization of the model of Ref.~\cite{Nonaka:2004pg}. We aim at gaining a comprehensive picture by considering a large number of trajectories also on the crossover side.  In fact we find that, depending on the parameterization, the effect may be more pronounced on the crossover side (e.g. top-left panel of Fig.\ \ref{fig:critical_lensing}). 

Here, we also differ from previous works by giving a quantitative thermodynamic argument for why this phenomenon applies for any dynamical system in which the EoS is physically relevant and the evolved densities take the system through the critical region (as was discussed in Sec.\ \ref{sec:criticallensing}).

\section{Results: Out-of-equilibrium}

\subsection{Critical lensing} \label{sec:hydroTraj}

In Sec.\ \ref{sec:criticallensing} we discussed the critical lensing effect in equilibrium.  The natural question is whether this effect can survive when the system is potentially far-from-equilibrium. At large $\mu_B$, the QGP evolution is influenced by multiple transport coefficients such as shear and bulk viscosity, as well as by conserved charge ($BSQ$) diffusion.  Currently, the far-from-equilibrium initial conditions at the beam energy scan are not known. Thus, it is hard to know how much guidance one can receive from equilibrium trajectories. For this reason, in this section we explore the possibility of an out-of-equilibrium critical lensing effect. 

\begin{figure}
    \centering
        \includegraphics[width=\linewidth]{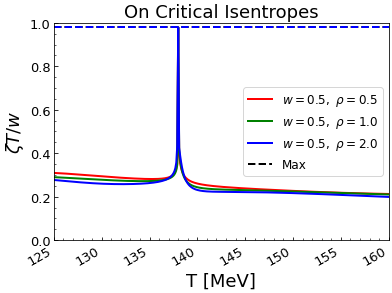}
    \caption{$\zeta T/w$ along isentropes that pass through the critical point for the three EoS shown in Fig.\ \ref{fig:traj}.   }
    \label{fig:bulkCP}
\end{figure}
\begin{figure*}
    \centering
    \begin{tabular}{ccc}
        \includegraphics[width=0.33\linewidth]{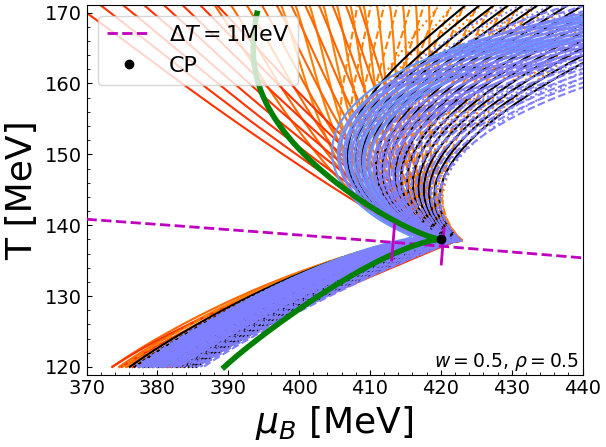}& 
        \includegraphics[width=0.33\linewidth]{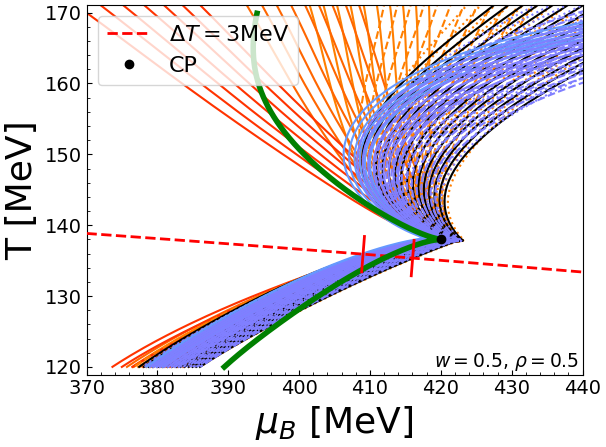}& 
        \includegraphics[width=0.33\linewidth]{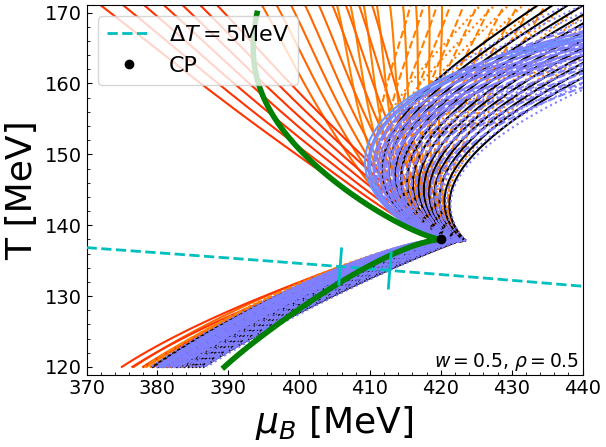}
        \\
        \includegraphics[width=0.33\linewidth]{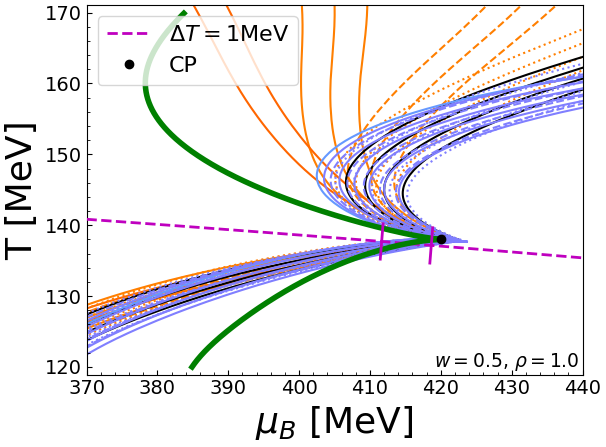}& 
        \includegraphics[width=0.33\linewidth]{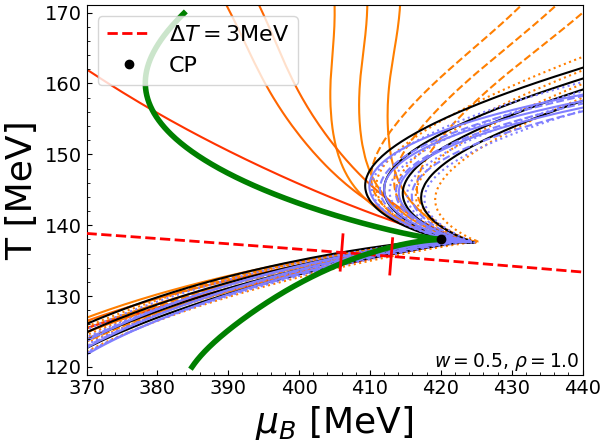}&
        \includegraphics[width=0.33\linewidth]{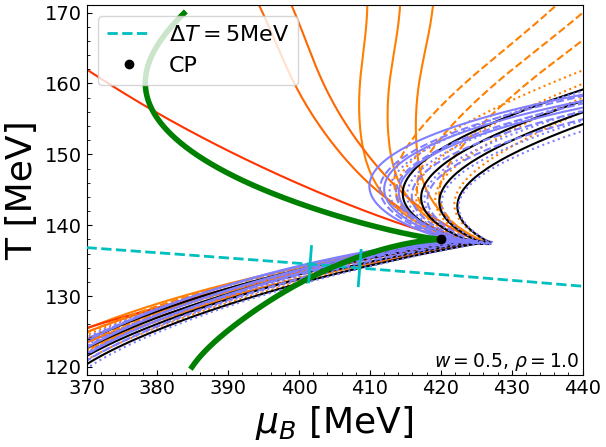}
         \\
         \includegraphics[width=0.33\linewidth]{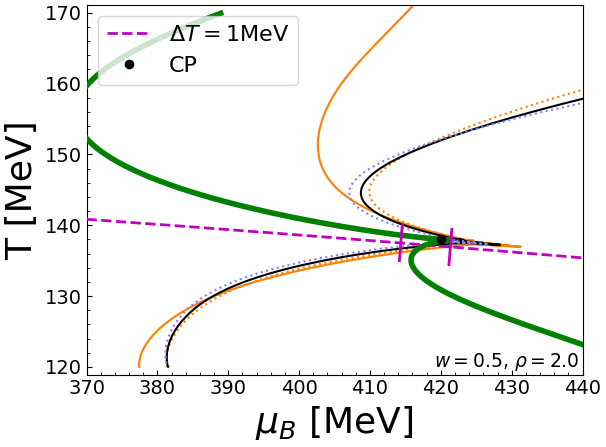}&
         \includegraphics[width=0.33\linewidth]{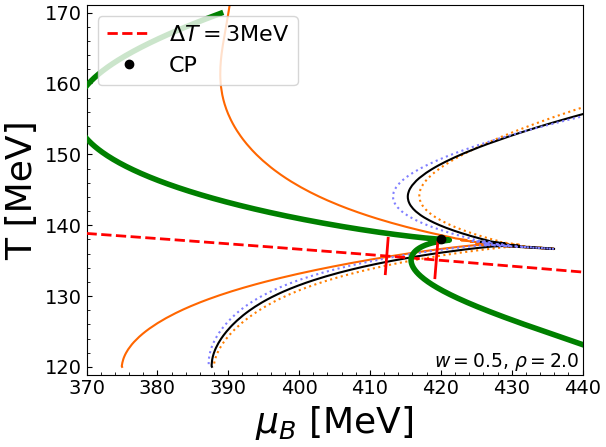}&
         \includegraphics[width=0.33\linewidth]{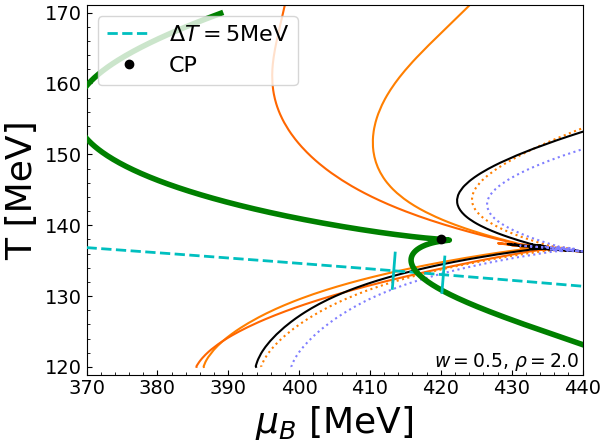}
    \end{tabular}
    \caption{Hydrodynamic trajectories for three different EoS with fixed $w=0.5$ and $\rho=\left\{0.5,1.0,2.0\right\}$ (top to bottom). Each column shows trajectories that pass within $\pm 3.5$ MeV plane from the critical isentrope, measured on a line parallel to the transition line, and shifted downwards by $\Delta T=\left\{1,3,5\right\}$ MeV (left to right). An extremely large number of initial conditions were run for each EoS, and only the trajectories that pass through our freeze-out window are shown.}
    \label{fig:traj}
\end{figure*}

All our hydrodynamic simulations use the same $\eta T/w (T,\mu_B)$, the only variability coming from the choice of equation of state, which in turn affects the bulk viscosity. The effect of the equation of state on $\zeta T/w$ is twofold:
\begin{enumerate}
    \item A minimum in $c_s^2$ appears at the critical point\footnote{At the time of finishing this paper, an orthogonal study on out-of-equilibrium $c_s^2$ at a critical point in holography was released \cite{Cartwright:2022hlg} } (see Fig.\ \ref{fig:1eos}), which in turn generates a peak in $\zeta T/w$; 
    \item The bulk viscosity scales with $\xi^3$ near the critical point, which further enhances $\zeta T/w$, see Eq.\ (\ref{eqn:zetaCS}).
 \end{enumerate}
Because of these two separate contributions, one anticipates a large enhancement in $\zeta T/w$ near the critical point. We show in Fig.\ \ref{fig:bulkCP} the bulk viscosity along the critical isentrope, for the three parameter choices in Fig.\ \ref{fig:traj}. Exactly at the critical point, universality forces the peaks to be identical. However, farther away from it, sub-leading contributions are such that a slightly larger $\zeta T/w$ is realized when $\rho$ is smaller, i.e. when the critical region extends more along the $T$-direction.

Next, we consider three of the parameter choices shown in Figs.\ref{fig:1eos},\ref{fig:critical_lensing}, namely $w=0.5$ with $\rho=0.5,1,2$ (top row in both figures). This is because, as we saw, the largest effect is given by the direction in which the critical region extends. This way, we can study three cases where some lensing is observed, but the cardinal orientation of the critical region varies from $T$ to $\mu_B$. 

In these three cases, we are able to investigate out-of-equilibrium effects with our simple hydrodynamic model. As discussed in Sec.\ \ref{hydro_Setup}, we run the simulations with a variety of initial conditions, in order to find trajectories that cross through a certain freeze-out window. As defined in Sec.\ \ref{hydro_Setup}, this window corresponds to $3.5$ MeV on either side of the critical isentrope, measured on a line parallel to the QCD transition line, and shifted downwards by $\Delta T=1,3,5 \MeV$ in order to account for some uncertainty in the freeze-out temperature, as mentioned earlier. 

Our initial conditions consist of an initial energy density, baryon density, shear stress, and bulk pressure, i.e. $\left\{\varepsilon,n_B,\pi_\eta^\eta,\Pi\right\}_0$. The equation of state maps trajectories in $\{ \varepsilon,n_B\}$ to trajectories in $\{T, \mu_B\}$. We initialize the baryon density $n_B$ with values ranging between $n_{B_0} = [0.4,1.0]$  fm$^{-3}$, with steps of $0.01$fm$^{-3}$, and the energy density is kept at $\varepsilon_0 = 1.5$ GeV/fm$^{-3}$. We initialize the dimensionless quantity $\pi_\eta^\eta/(\varepsilon+p)$ with values ranging between $\pi_\eta^\eta/(\varepsilon+p)= [-0.5,0.5]$ with steps of $0.2$. Similarly, we initialize the dimensionless quantity $\Pi/(\varepsilon+p)$ with values ranging between $\Pi/(\varepsilon+p)= [-0.5,0.5]$ with steps of $0.2$. Combining all choices independently, we have a multidimensional grid of $61\times6\times 6\times 1=2196$ initial conditions for each equation of state. These initial conditions are chosen such that they allow us to scan the entire $\left\{T,\mu_B\right\}$ plane available within the limitations of the BEST EoS (the BEST EoS can become acausal/thermodynamically unstable beyond $\mu_B\gtrsim 600$ MeV, due to the limited number of susceptibilities $\chi_n^B$ available from lattice simulations). 

We show our hydrodynamic trajectories in Fig. \ \ref{fig:traj}, for the three values of $\rho$ (top to bottom), and the three values of $\Delta T$ (left to right). The freeze-out line is shown as a dashed curve shifted downwards by $\Delta T$ from the transition line, and the freeze-out window is denoted by two solid lines perpendicular to the freeze-out line. Only trajectories that pass through such freeze-out window for a specific $\Delta T$ are shown.

Notably, from Fig.\ \ref{fig:traj} it seems evident that the value of $\rho$ is much more important than that of $\Delta T$. When $\rho$ is smaller, a  significantly larger number of trajectories pass within the freeze-out window, regardless of the definition of freeze-out temperature.  
Comparing with Fig.\ \ref{fig:critical_lensing}, we find a consistent picture. The same effect seen in equilibrium survives even when far-from equilibrium initial conditions are used: essentially, we are observing something we can call \emph{dynamical} critical lensing.

This \textit{dynamical} critical lensing provides an exciting possibility. Even though heavy-ion collisions may initially be far-from equilibrium, given a critical point with a critical region as we have just described, an attractor may exist that pushes their evolution trajectories towards the critical point.  It would be extremely interesting to explore this effect in more realistic hydrodynamic simulations, in 2+1D or 3+1D, since higher dimensions would allow for the incorporation of $BSQ$ diffusion, flow effects, and the rapidity dependence of baryon density. 

There is, of course, the possibility that the scenario realized in Nature is the opposite, namely that the critical region extends mostly along the $\mu_B$ direction. In such a case, very few trajectories would converge towards the critical point, making its detection much more challenging. We have checked the qualitative features we just discussed on many more parameter choices than we could present, and can confirm the general trend that critical lensing is enhanced when the critical region extends along the temperature direction.

\subsection{Deviations from Isentropes and Entropy Production}

\begin{figure*}
    \centering
    \begin{tabular}{ccc}
        \includegraphics[width=0.33\linewidth]{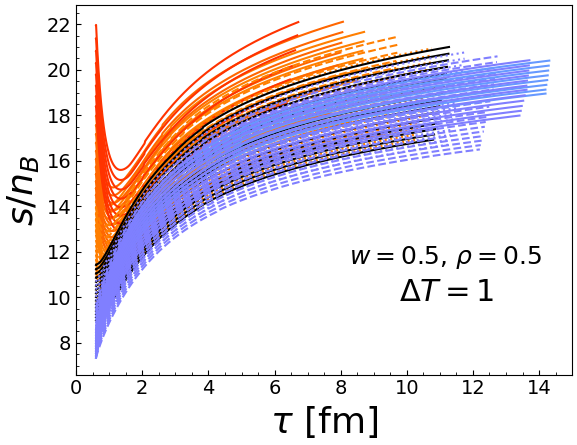}& 
        \includegraphics[width=0.33\linewidth]{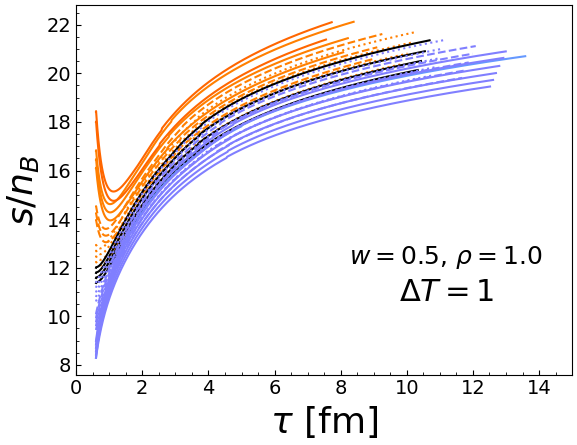}& 
        \includegraphics[width=0.33\linewidth]{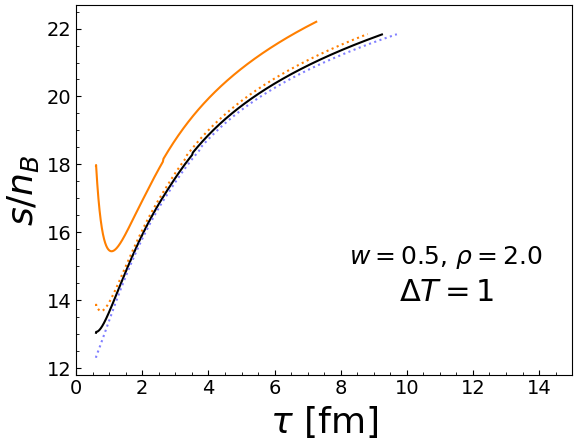}
    \end{tabular}
    \caption{Thermal entropy over baryon number ($s/n_B$) for the same hydrodynamic trajectories shown in the left column of Fig.\ \ref{fig:traj} ($w=0.5$ fixed and $\rho=\left\{0.5,1.0,2.0\right\}$, $\Delta T=1$ MeV).}
    \label{fig:snb}
\end{figure*}

We saw in the last section that the hydrodynamic trajectories never seem to converge to the isentropic ones. One might ask why that is. What would happen if the initial conditions were chosen at equilibrium? The trajectories would not match the isentropes even in this case, because the QGP (and thus our setup) has non-vanishing shear and bulk viscosities, so that the magnitude of the shear stress tensor and bulk pressure grow over time, i.e. entropy is produced. The evolution trajectories would resemble the isentropes only if the initial conditions were chosen at equilibrium, \textit{and} the viscosities vanished (i.e. ideal hydrodynamics in equilibrium).  Moreover, because we do know that $\eta T/w$ may grow further at large $\mu_B$, and that the bulk viscosity is sensitive to critical scaling,  it is all more important to employ relativistic viscous hydrodynamic simulations.

When viscosity is included within a hydrodynamic framework, entropy is no longer conserved, but rather it is produced. The amount of entropy production is dependent on how far-from-equilibrium the fluid is throughout its evolution.  In  heavy-ion collisions, it is often assumed that entropy production is small because
both $\eta T/w$ and $\zeta T/w$ are small. However, in short-lived systems that may begin far-from-equilibrium, that might not be the case. Additionally, it is not guaranteed that $\eta T/w$ and $\zeta T/w$ are small also at large chemical potential. One should then consider the possibility that a large amount of entropy is produced.  

Calculating the amount of entropy production in hydrodynamic simulations is quite challenging, because it receives contributions from both thermal entropy and out-of-equilibrium entropy. In our setup, we cannot estimate the out-of-equilibrium entropy, and can only calculate the thermal entropy from the equation of state. This means that our results can demonstrate that thermal entropy is produced, but \emph{additional} contributions from out-of-equilibrium entropy might exist, which we are unable to track. We should emphasize that the semi-positive-definiteness of the entropy change applies to the \textit{total} entropy, thus it is possible that this change is negative when the thermal entropy alone is considered.

With this caveat in mind, we show in Fig.\ \ref{fig:snb} the thermal contribution to the ratio $s/n_B$, for the same paramemeterizations of the equation of state shown in Fig.\ \ref{fig:traj}, along the trajectories obtained with $\Delta T=1$ MeV.  We find that an enormous amount of entropy is produced from early times until freeze-out, which explains the substantial difference between equilibrium and out-of-equilibrium trajectories we have previously observed.

We also observe that, for $\tau \lesssim 2 \fm$, some trajectories move downwards, which implies a negative change in thermal entropy. This is not necessarily an issue, because -- as already mentioned -- the semi-positive-definiteness of entropy applies to the \textit{total} entropy. However, it is also possible that some trajectories do violate certain causality conditions (see. \cite{Bemfica:2020xym,Plumberg:2021bme,Chiu:2021muk}). At this time, we have only checked the weak energy \cite{Janik:2005zt, Dore:2020jye} condition, which is not as stringent as the nonlinear causality constraints. Another possibility it that, in these regimes, the system exhibits non-hydrodynamic behavior such as cavitation \cite{Byres:2019xld,Denicol:2015bpa,Rajagopal:2009yw}. Should that be the case, it is possible to extend the model to account for these effects in a way that guarantees stability \cite{Torrieri:2007fb,Torrieri:2008ip}. However, this is beyond the scope of this work.

The trajectories that experience this behavior are shown in red and orange colors, which indicate initial conditions with $\Pi>0$  and $\pi^\eta_\eta<0$, atypical for heavy-ion collisions. 
However, you do achieve $\Pi>0$ in heavy-ion collision simulations when you match a conformal initial condition to the non-conformal hydrodynamic simulations due to the mismatch in EoS \cite{NunesdaSilva:2020bfs}. 
In contrast, the purple, blue, and turquoise lines correspond to values typically found in heavy-ion collisions.

Previous attempts have been made to compare lines of $s/n_B$ from heavy-ion collisions (from ideal hydrodynamics) to neutron star mergers \cite{Most:2022wgo}. Our findings suggest that very large deviations should be anticipated due to entropy production. Thus, one truly requires a solid understanding of the dissipative effects at large densities in order to make a comparison between these two systems.  Moreover, because we cannot take into account $BSQ$ diffusion effects in our framework, we anticipate even larger deviations from isentropes would occur when such effects are incorporated in full 3+1 relativistic viscous hydrodynamic simulations. 

\subsection{Out-of-equilibrium effects on kurtosis}
\begin{figure*}
    \centering
    \begin{tabular}{ccc}
        \includegraphics[width=0.33\linewidth]{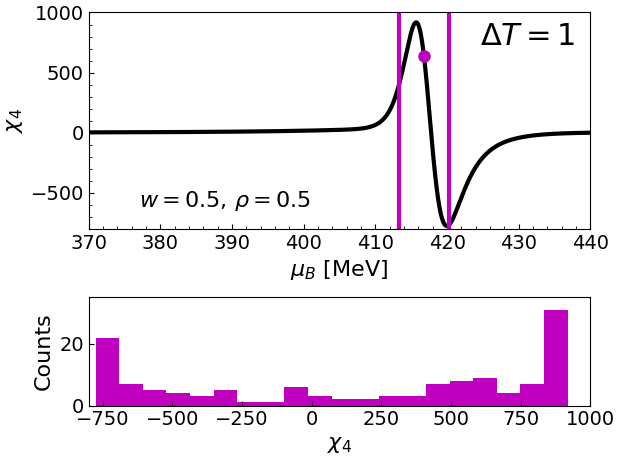}& 
        \includegraphics[width=0.33\linewidth]{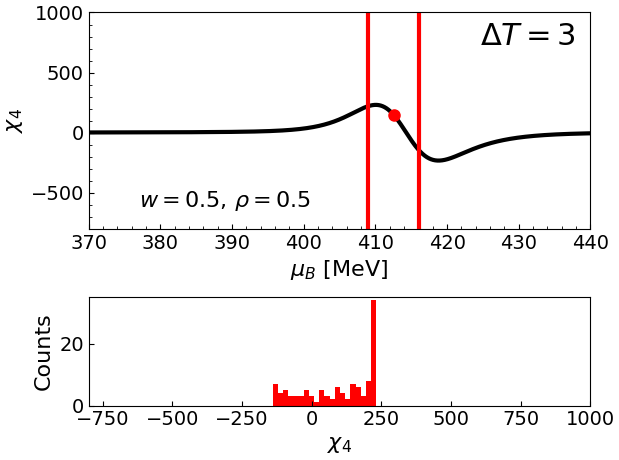}& 
        \includegraphics[width=0.33\linewidth]{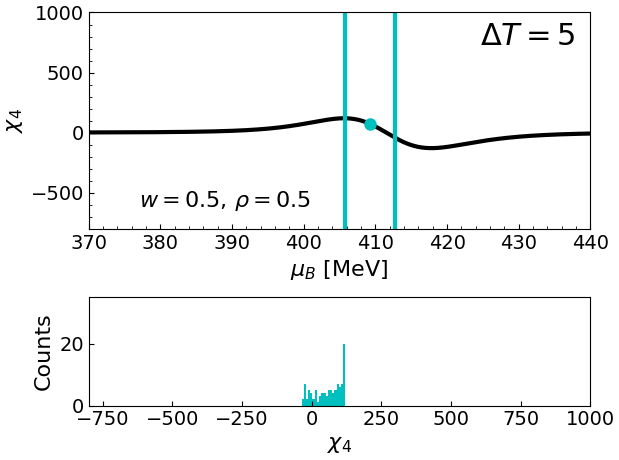}
        \\
        \includegraphics[width=0.33\linewidth]{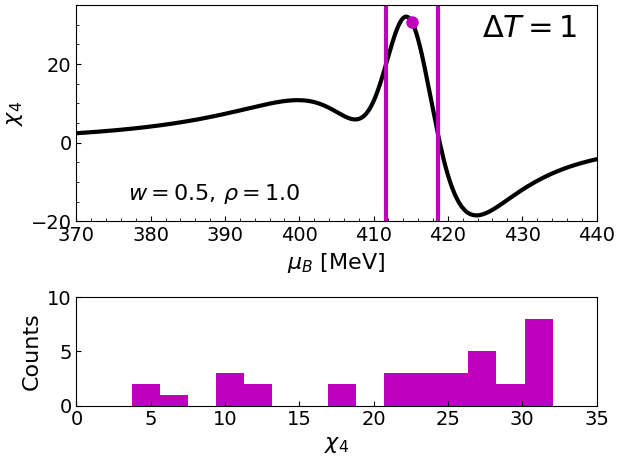}& 
        \includegraphics[width=0.33\linewidth]{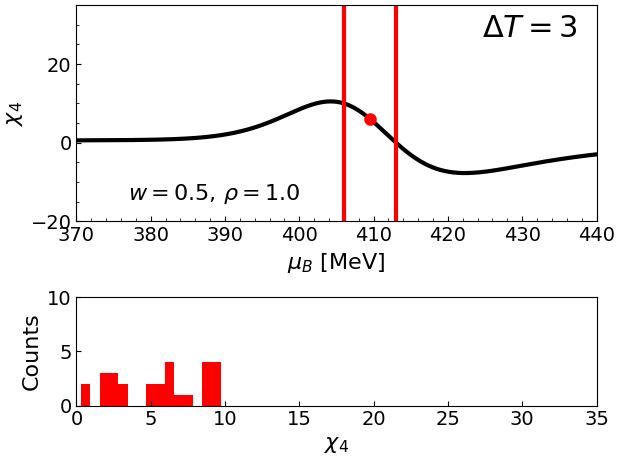}&
        \includegraphics[width=0.33\linewidth]{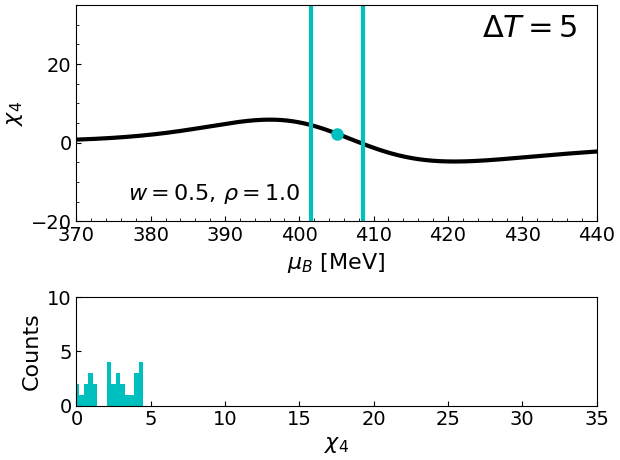}
         \\
         \includegraphics[width=0.33\linewidth]{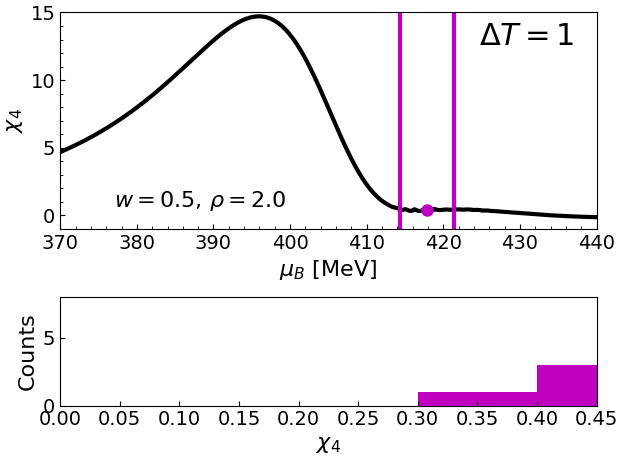}&
         \includegraphics[width=0.33\linewidth]{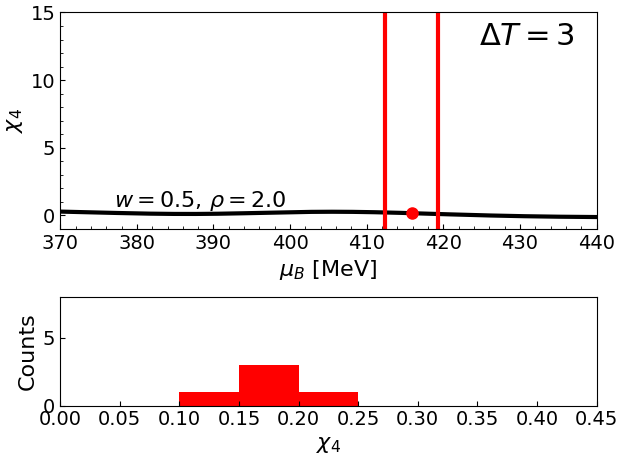}&
         \includegraphics[width=0.33\linewidth]{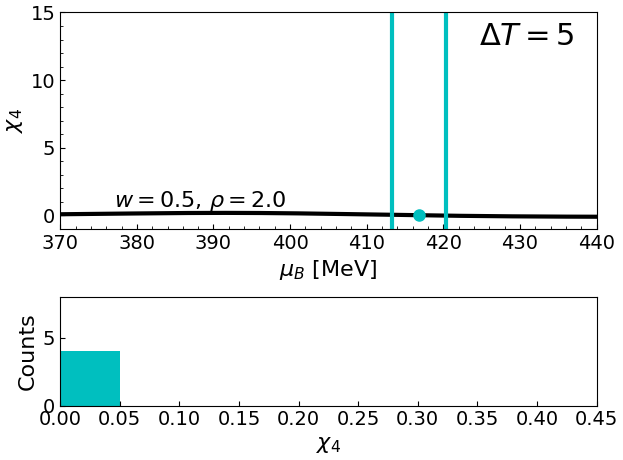}
    \end{tabular}
    \caption{Fourth baryon susceptibility as a function of the chemical potential for the same parametrizations of the equation of state as in Fig. \ref{fig:traj}, with fixed $w=0.5$ and $\rho=\left\{0.5,1.0,2.0\right\}$ (top to bottom), and $\Delta T=\left\{1,3,5\right\}$ MeV (left to right). The dot on each line is where the isentrope intersects the freeze-out line, and the vertical line signal the freeze-out window. Below each plot is the histogram for the values of $\chi_4$ obtained from out-of-equilibrium trajectories. }

    \label{fig:kurtosis_outofEQ}
\end{figure*}

Early works argued that, besides the peak in the net-baryon number kurtosis, a dip was to be expected at larger collision energies, as a sign of the QCD critical point \cite{Stephanov:2011pb,Pradeep:2019ccv}. However, not all effective models predicting a critical point exhibit such a dip (see e.g. \cite{Critelli:2017oub,Grefa:2021qvt}, where $\chi_4$ monotonically increases approaching the critical point). Furthermore, it was recently discovered that, when including sub-leading effects due to the mapping between Ising model and QCD phase diagram (not considered in the earlier works), such dip appears not to be a robust feature of the dependence of the kurtosis on the collision energy \cite{Mroczek:2020rpm}.

We have seen that non-equilibrium effects play a significant role in the evolution of the system, and in this section we will investigate how these effects influence the kurtosis, by looking at all different trajectories that fall within our previously defined freeze-out windows. In Fig.\ \ref{fig:kurtosis_outofEQ}, we show $\chi_4$ as a function of $\mu_B$. 

We consider the same three EoS, and the same trajectories shown in Fig. \ref{fig:traj}, with the same layout: $w=0.5$ always, $\rho=0.5,1,2$ (top to bottom), $\Delta T=1,3,5 \MeV$ (left to right). We highlight the freeze-out windows with solid, colored lines, and the point where the isentrope intersects the freeze-out line with a colored dot. Below each of these plots, we show the histogram of the outcomes of $\chi_4$ from all the trajectories. The resulting measured $\chi_4$ would be a convolution of such histograms.
Though not extremely apparent, a couple of trends can be observed from these plots. As we already knew, the total number of entries decreases when $\rho$ increases, due to the reduced lensing effect on the trajectories. On the other hand, when increasing $\Delta T$, the peakedness of the distribution decreases, because having a later freeze-out allows for selecting trajectories that span a larger set of chemical potentials. 
Overall, the resulting $\chi_4$ is predominantly positive, which is encouraging in view of actual measurements, which, like in our simplified setup, will be forced to effectively ``integrate'' over a range of chemical potentials, due to the finite width of rapidity bins in the analysis.

\begin{figure}
    \centering
    \includegraphics[width=\linewidth]{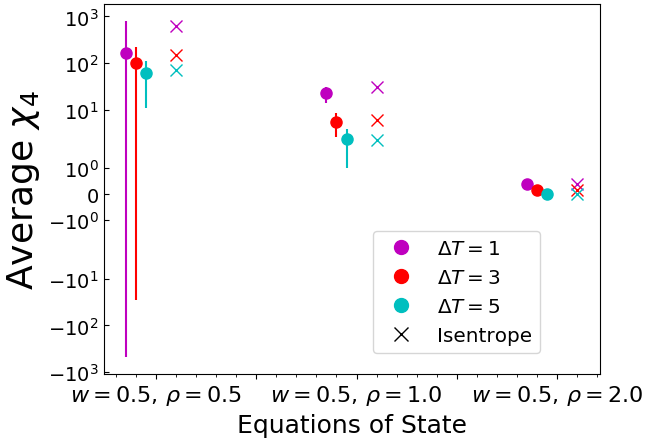}
    \caption{Average $\chi_4$ across all hydrodynamic trajectories shown in Fig.\ \ref{fig:kurtosis_outofEQ} for various temperatures differences, $\Delta T$, between the hadronization and freeze-out temperatures.}
    \label{fig:avg_chi4}
\end{figure}

In Fig.\ \ref{fig:avg_chi4} we show an ``averaged" $\chi_4$, obtained by integrating over the probability distributions shown in Fig.\ \ref{fig:kurtosis_outofEQ}.  We show this for the same $\rho$, $w$, $\Delta T$ combinations as previously shown. The value of the isentrope that passes exactly through the critical point is shown in $X$ whereas the average $\chi_4$ over all trajectories at that $\Delta T$ is shown in the filled in colored circle. In addition, one can find one standard deviation away from the average $\chi_4$ with the lines.  We generally find that, indeed, small $\rho$ and $w$ lead to a large, positive $\chi_4$. In contrast, increasing $\rho$ significantly suppresses $\chi_4$.  For larger separations between the hadronization and freeze-out temperatures $\Delta T$, $\chi_4$ is somewhat suppressed, but this effect is significantly smaller than that of the choice of $\rho$, $w$. We have also checked effects of the freeze-out window choice on the $\chi_4$.  Even with a larger window we still see the same effect with preferences for trajectories to hit low and high points in the $\chi_4$. In fact, if the window includes both the peak and dip, the effect of trajectories being pulled towards the max or min value of $\chi_4$ is even more robust. Obviously, though, an extremely large window will allow for more fluctuations in $\chi_4$ as well.

One complication that can arise from a larger window is capturing both the peak and dip of $\chi_4$ symmetrically. In this case, it's possible for the average $\chi_4$ seen by the trajectories to go to zero. In line with this, what is interesting to note, is that for small $\rho$ and $w$ the deviation between the equilibrium $\chi_4$ (along the isentrope) versus the average $\chi_4$ is larger.  This is not surprising because small $\rho$ and $w$ also experience more out-of-equilibrium critical lensing effects. Thus, this suppression of average $\chi_4$ is a consequence of critical lensing.  However, for these smaller freeze-out windows, even with the out-of-equilibrium smearing of the average $\chi_4$ for small $\rho$ and $w$ the central values remain consistently larger than $w=0.5, \rho=1.0$.  If we then look at 1 standard deviation away from the central value it is clear that there's a skewed towards larger values of $\chi_4$ but there's some change of extremely small values (or negative values) as well. Already for $w=0.5, \rho=1.0$ the critical lensing effect is small enough that there is almost no difference between the equilibrium value of $\chi_4$ and the out-of-equilibrium average $\chi_4$, even when one considers 1 standard deviation from the mean. In contrast, for $w=0.5, \rho=0.5$, there is a large standard deviation which is a consequence of the rapid change in $\chi_4$ in the freeze-out window.

It is worth mentioning a caveat in comparing this averaged $\chi_4$ to what is actually measured in experiment. In this work, we have access to each individual ``event" and are therefore able to select on the specific trajectories that pass through a given freeze-out window. In experiment this is not possible. Instead, the experimentally measured freeze-out temperature and chemical potential are extracted for a given beam energy over a large ensemble of events. Thus, subtle difference exist due the the simplicity of our toy model.  However, even with these differences, works such as this one and Ref. \cite{Dore:2020jye} clearly indicate a non-trivial relation between the initial state and final freeze-out. In fact, recent studies show that these far-from-equilibrium effects may be enhanced for increased chemical potential \cite{Rougemont:2022piu}. Motivated by these results, we will directly connect to experimental data using realistic 2+1 and 3+1 relativistic  hydrodynamic viscous models in future work.

\section{Conclusions}

In this work we explored the effect of different parametrizations of the BEST Collaboration equation of state, which affect the shape and size of the critical region around the QCD critical point, on the net-baryon number kurtosis, and on critical lensing. We found that the direction along which the critical region extends is also a relevant factor, besides its size. The lensing effect was observed both in equilibrium, as well as in out-of-equilibrium simulations. In both cases, critical regions that further extend along the $T$-direction were shown to induce the largest critical lensing effect, even when the system was initialized far-from-equilibrium.  

Because of this, many more evolution trajectories passed through the vicinity of the critical point, which would make its detection more likely in an experimental setting.

While in ideal hydrodynamics entropy is conserved, meaning that isentropes serve as good proxies for the hydrodynamic trajectories through the QCD phase diagram, the presence of viscosity induces a generous entropy production, which makes isentropes a poor guide for realistic scenarios. This was found to be the case regardless of the equation of state used. As confirmation, we showed clear evidence for the large thermal production of entropy during the whole system's evolution. Additionally, ours is likely a conservative estimate, considering that we could not estimate the contribution from out-of-equilibrium entropy production, and that additional effects (e.g. $BSQ$ diffusion) are expected to play a role, especially in higher dimensions.   

Finally, we investigated the spread in the kurtosis at freeze-out, using our hydrodynamic trajectories with different equations of state, taking into account the uncertainty on the freeze-out temperature. We found that the critical lensing induces a non-trivial distribution in $\chi_4$ at freeze-out, which becomes more evident, the closer the freeze-out point is from the transition line. 

This is quite a non-trivial effect, because it would have a significant impact on the experimentally measured kurtosis. 

In addition, a critical region extending along the $T$-direction produces much larger fluctuations in $\chi_4$, such that large positive or large negative values of $\chi_4$ are possible at freeze-out (this is due to the sharpness in the peak of $\chi_4$ and non-monotonic behavior at $\mu_B > \mu_{BC}$).   Critical regions that extend further along the $T$-direction, which produce a stronger lensing effect, were also previously found to be preferred by lattice results at $\mu_B = 0$ \cite{Mroczek:2022oga}. In contrast, for a critical region extending further along the $\mu_B$-direction, $\chi_4$ is significantly smaller and less likely to present a clear signal.  However, even with large fluctuations for critical regions along the $T$-axis, the average $\chi_4$ ends up being large and clearly positive, whereas it is clear that critical regions along the $\mu_B$ direction have orders of magnitude smaller average $\chi_4$. We find that the difference between the hadronization temperature and freeze-out temperature plays a smaller role than the difference in the EoS themselves.

To our knowledge, this is the first study wherein different type of critical regions were compared, while coupling to full viscous hydrodynamics. Certainly, a number of effects remain to be explored. The most obvious next step is to move to higher-dimensions in the equations of motion, i.e. with 1+1D  \cite{Fotakis:2019nbq} or 3+1D setups \cite{Denicol:2018wdp,Du:2019obx,Schafer:2021csj}. Already at 1+1D, diffusion can be considered, which is expected to be suppressed at the critical point \cite{Du:2021zqz}. Furthermore, a non-trivial coupling between $BSQ$ conserved currents exists \cite{Greif:2017byw}, and diffusion currents also couple to shear and bulk viscosity at the level of the equations of motion \cite{Denicol:2012cn,Monnai:2012jc,Fotakis:2022usk}. It remains to be seen whether these effects are even further enhanced in more realistic simulations. At this point, we are still quite far from studies that can make direct comparisons to experimental data, because this would require a freeze-out procedure that conserves $BSQ$ charges followed by hadronic transport \cite{Oliinychenko:2019zfk,Oliinychenko:2020cmr}. Thus, we cannot comment e.g., on the effects of kinematic cuts at this time. However, it has been shown that the anti-proton-to-proton ratio $\bar{p}/p$ may be sensitive to deformations in the trajectories \cite{Asakawa:2008ti}. It is unclear how strong out-of-equilibrium effects at freeze-out would change this, since these corrections may affect this ratio. 
Finally, memory effects may play a significant role in these types of simulations \cite{Mukherjee:2016kyu}, which would be interesting to study in a future work. 

{\bf \emph{Acknowledgements} -- }

The authors would like to thank Mauricio Hippert for insightful discussion about that nature of the critical lensing effect. D.M. is supported by the National Science Foundation Graduate Research Fellowship Program under Grant No. DGE – 1746047 and the University of Illinois at Urbana-Champaign Sloan Graduate Fellowship. T.D. and D.M. acknowledge support from the ICASU Graduate Fellowship. J.N.H. acknowledges financial support by the US-DOE Nuclear Science Grant No. DESC0020633. Y.Y. is supported by the U.S.~Department of Energy under Contract No.~DE-FG02-93ER-40762. J.M.K. is supported by an Ascending Postdoctoral Scholar Fellowship from the National Science Foundation under Award No. 2138063. 
C.R. acknowledges financial support
by the National Science Foundation under grant no. PHY-1654219. I.L. acknowledges support by the National Science Foundation and Department of Defense under Grant PHY-1950744. Any opinions, findings, and conclusions or recommendations expressed in this material are those of the author(s) and do not necessarily reflect the views of the National Science Foundation or Department of Defense.

\bibliography{inspire, NOTinspire}
\appendix
\section{Critical Scaling of Thermodynamic Variables}\label{sec:appendixA}
We can rearrange the map between Ising and QCD variables to write
\beq
h(T,\mu_B) = \frac{\tan \left(\alpha _1\right) \left(\mu _B-\mu_{BC}\right)+(T-T_C)}{T_C w( \sin \left(\alpha _2\right)-   \cos \left(\alpha _2\right) \tan \left(\alpha _1\right))}
\eeq
\beq
r(T,\mu_B) = \frac{\tan \left(\alpha _2\right) \left(\mu _B-\mu_{BC}\right)+(T-T_C)}{\rho  w T_C (\sin \left(\alpha
   _1\right)- \cos \left(\alpha _1\right) \tan \left(\alpha _2\right))},
\eeq
and define the differential operations
\beq\label{eq:operatorT}
\partial_T = h_T\partial_h + r_T\partial_r 
\eeq
\beq\label{eq:operatormuB}
\partial_{\mu_B} = h_{\mu_B}\partial_h + r_{\mu_B}\partial_r,
\eeq
where the subscripts correspond to partial derivatives (e.g. $h_T = \frac{\partial h}{\partial T}|_{\mu_B}$) and
\beq
\partial_h \sim \dfrac{r^{1 - \beta\delta}}{\beta\delta}\partial_r.
\eeq
We obtain the critical scaling of different thermodynamic variables by applying the operations in Eqs. (\ref{eq:operatorT}) and (\ref{eq:operatormuB}) to the pressure as defined in Eq. \ref{eq:criticalp}, 
%entropy
\begin{align}
s \sim \partial_{T} P^{crit} = & S_0\frac{r^{\beta }}{T_C w \left( \sin \left(\alpha _2\right)-\cos \left(\alpha _2\right) \tan
   \left(\alpha _1\right)\right)} \nonumber \\
   & +S_1\frac{ r^{\beta  \delta +\beta -1}}{T_C \rho  w \left(\sin \left(\alpha _1\right)-\cos \left(\alpha _1\right) \tan \left(\alpha
   _2\right)\right)}    
\end{align}
%Number density
\begin{align}
n_{B} \sim \partial_{\mu_B} P^{crit} = & N_0\frac{ \tan \left(\alpha _1\right)r^{\beta }}{ T_C w \left(\sin \left(\alpha
   _2\right)-\cos \left(\alpha _2\right) \tan \left(\alpha _1\right)\right)} \nonumber \\
   & + N_1\frac{ \tan \left(\alpha _2\right)  r^{\beta  \delta +\beta -1}}{T_C \rho  w \left( \sin \left(\alpha _1\right)-\cos \left(\alpha
   _1\right) \tan \left(\alpha _2\right)\right)},
\end{align}
where $S_i$ and $N_i$ are constants which depend only on $\beta$ and $\delta$. The expressions for second-order derivatives are significantly longer and are shown below only up to leading order in $r$,
%dsdT
\begin{align}
\left(\dfrac{\partial s}{\partial T}\right)_{\mu_B} & \sim (\partial_{T})^2 P^{crit} \nonumber\\
& \sim \frac{1}{r^{\beta\delta-\beta} T_C^2 w\left( \sin \left(\alpha _2\right)- \cos \left(\alpha _2\right) \tan \left(\alpha_1\right)\right)^2},
\end{align}
%chi2
\begin{align}
\chi_2^B & \sim (\partial_{\mu_B})^2 P^{crit} \sim \frac{\sin ^2\left(\alpha _1\right) \csc ^2\left(\alpha _1-\alpha _2\right)}{r^{\beta\delta -\beta} T_C^2 w^2},
\end{align}
%dsdmu and dndT
\begin{align}
\left(\dfrac{\partial s}{\partial \mu_B}\right)_T  = \left(\dfrac{\partial n_B}{\partial T}\right)_{\mu_B} & \sim \partial_{T}\partial_{\mu_B} P^{crit} \nonumber\\
& \sim \frac{\sin \left(2 \alpha _1\right) \csc ^2\left(\alpha _1-\alpha _2\right)}{r^{\beta\delta - \beta} T_C^2 w^2}.
\end{align}
We obtain the scaling behavior of the $T$ and $\mu_B$ separation between isentropes by substituting the leading terms for each quantity into Eqs. (\ref{eq:muBseparation}) and (\ref{eq:Tseparation}), resulting in Eq. (\ref{eq:scalingofisentropeseparation}).

\section{Scaling of the Separation Between Isentropes as a Function of Equation of State Parameters}\label{sec:appendixB}
At the crossover line, $T$ is a function of $\mu_B$, as specified by Eq. (\ref{eq:trline}), so Eq. (\ref{eq:muBseparation}) becomes
\begin{align}
    \dfrac{d(s/n)}{d\mu_B} &= \frac{1}{n}\left( \dfrac{\partial s}{\partial T}\dfrac{\partial T}{\partial \mu_B} + \frac{\partial s}{\partial \mu_B} \right) \\ \nonumber
    & \, \, \, \, - \frac{s}{n^2}\left(  \frac{\partial n}{\partial T}\dfrac{\partial T}{\partial \mu_B} + \frac{\partial n}{\partial \mu_B} 
    \right)  \, \,.
\end{align}

Near the critical point, we can write
\begin{eqnarray}
    \dfrac{d(s/n)}{d\mu_B}&\sim &(\partial_{\mu_B} P^{crit})^{-1}\left(\partial_T P^{crit} \dfrac{\partial T}{\partial \mu_B}  + \partial_{\mu_B}\partial_T P^{crit} \right) \\ \nonumber 
     &-&\frac{\partial_T P^{crit}}{(\partial_{\mu_B} P^{crit})^2}\left(  \partial_T\partial_{\mu_B} P^{crit} \dfrac{\partial T}{\partial \mu_B} + \partial_{\mu_B}\partial_{\mu_B}P^{crit}\right),
\end{eqnarray}
and by using the operations defined in Eqs. (\ref{eq:operatorT}, \ref{eq:operatormuB}), obtain the behavior of $d(s/n)/d\mu_B$ along the crossover line, near the critical point, as a function of the Ising variable $r$ and the EoS input parameters,
\begin{widetext}
\beq
\dfrac{d(s/n)}{d\mu_B} \sim \frac{(\beta \delta -1) \left(2 \beta \delta  \kappa_2  \mu_B  \cos \left(\alpha_2\right) r^{\beta  \delta }+\beta  \delta  T_0 \sin \left(\alpha _2\right) r^{\beta  \delta
  }-2 \kappa_2  \mu_B  \rho  r \cos \left(\alpha _1\right)-\rho  r T_0 \sin \left(\alpha _1\right)\right)}{T_0 T_C w \left(\rho  r \sin \left(\alpha _1\right)-\beta
    \delta  \sin \left(\alpha _2\right) r^{\beta  \delta }\right){}^2}.
\eeq
\end{widetext}

Using both the approximation for the exact values of the 3D Ising exponents, $\beta = 1/3$, $\delta = 5$, and the mean-field values $\beta = 1/2$, $\delta = 3$, the leading terms are the same up to an overall constant $A^*$
\beq
\dfrac{d(s/n)}{d\mu_B} \sim A^*\frac{\csc \left(\alpha _1\right) \left(2 \kappa_2  \mu_B  \cot \left(\alpha _1\right)+T_0\right)}{ w \rho  r T_0 T_C} + \dots \quad .
\eeq

%\end{document}

\end{document}